\documentclass[aps,prb,twocolumn,superscriptaddress,amsmath,amssymb,floatfix, longbibliography,nofootinbib,notitlepage]{revtex4-2}
\usepackage{amsbsy}
\usepackage{amsfonts}
\usepackage{amsmath}
\usepackage{amstext}
\usepackage{amsthm}
\usepackage{amssymb}

\usepackage{graphicx}

\usepackage{epsfig}
\usepackage{epstopdf}
\usepackage{bm}
\usepackage{color}
\definecolor{prlblue}{rgb}{0.18,0.19,0.57}
\usepackage{multirow}
\usepackage[colorlinks]{hyperref}
\usepackage{cleveref}
\hypersetup{citecolor=prlblue, linkcolor=prlblue, filecolor=prlblue, urlcolor=prlblue}
\newcommand{\diag}[2]{\vcenter{\hbox{\includegraphics[height=#2]{#1}}}}
\newcommand{\figref}[1]{Fig.\,\ref{#1}}
\newcommand{\secref}[1]{Sec.\,\ref{#1}}
\newcommand{\eqnref}[1]{Eq.\,\eqref{#1}}
\newcommand{\appref}[1]{Appendix~\ref{#1}}
\renewcommand{\Im}{\mathop{\mathrm{Im}}}
\renewcommand{\Re}{\mathop{\mathrm{Re}}}

\begin{document}
\graphicspath{{figures/}}

\title{Crossover from Fermi Arc to Full Fermi Surface}

\author{Jia-Xin Zhang}
\affiliation{Institute for Advanced Study, Tsinghua University, Beijing 100084, China}
\author{Zheng-Yu Weng}
%\email{weng@mail.tsinghua.edu.cn}
\affiliation{Institute for Advanced Study, Tsinghua University, Beijing 100084, China}

\date{\today}
%%%%%%%%%%%%%%%%%%%%%%%%%%%%%%%%%%%%%%%%%%%%%%%%%%%%%%%%%%%%%%%%%%%%%%%%%%%%%%%%%%%%%%%%%%%%%%

\begin{abstract}
The Fermi surface as a contour of the gapless quasiparticle excitation in momentum space is studied based on a mean-field theory of the doped Mott insulator, where the underlying pseudogap phase %at low doping 
is characterized by a two-component resonating-valence-bond (RVB) order that vanishes in the overdoping at $\delta>\delta^*$. Here the quasiparticle emerges 
as a ``collective'' mode and a Fermi arc %structure 
%Barring superconducting phase coherence}, 
is naturally present in the pseudogap regime, while a full Fermi surface is recovered % the overdoping without RVB.  %Nevertheless, upon a careful examination, we show that 
at $\delta>\delta^*$. The area enclosed by the gapless quasiparticle contour still satisfies the Luttinger volume in both cases, and the ``Fermi arc'' at $\delta<\delta^*$ is actually due to a significant reduction of the spectral weight caused by a quasiparticle fractionalization %into more elementary particles 
in the antinodal region. %In particular, 
The endpoints of the Fermi arcs exhibit enhanced density of states or ``hotspots'', which can further give rise to a charge-density-wave-like quasiparticle interference pattern. 
%We also find that small Fermi-pockets violating the Luttinger volume can indeed emerge if the secondary doping-induced RVB order is suppressed alone, say, by strong perpendicular magnetic fields.
At the critical doping $\delta^*$, %beyond which the primary RVB order vanishes, 
the fractionalized spin excitations become gapless and incoherent which is signaled by a divergent specific heat. At $\delta>\delta^*$, the quasiparticle excitation restores the coherence over the full Fermi surface, but the fractionalization still persists at a higher energy/temperature which may be responsible for a strange metal behavior. Different mechanisms for the Fermi arc and experimental comparisons are briefly discussed.

\end{abstract}

\maketitle

\tableofcontents

%%%%%%%%%%%%%%%%%%%%%%%%%%%%%%%%%%%%%%%%%%%%%%%%%%%%%%%%%%%%%%%%%%%%%%%%%%%%%%%%%%%%%%%%%%%%%%
\section{Introduction}

%The fate of the quasiparticle excitation is an important characterization of the high-$T_c$ cuprate.
In the conventional BCS theory, quasiparticles as gapless excitations of a Fermi-liquid state will form the Cooper pairs and simultaneously experience superconducting (SC) condensation below $T_c$. However, a quasiparticle excitation in the ``normal state'' of an underdoped cuprate seems rather unconventional. As revealed by the angle-resolved photoemission spectroscopy (ARPES) experiments \cite{Shen.Damascelli.2003,Shen.Marshall.1995, Hinks.Norman.1998,Shen.Shen.2005kf5}, the gapless single-electron excitation forms \emph{incomplete} Fermi surface portions in the Brillouin zone, known as the ``Fermi arcs'' in the so-called pseudogap phase. Only in the overdoped cuprate can a full Fermi surface be restored above $T_c$ \cite{Campuzano.Kaminski.2003,Damascelli.Plat.2005,Campuzano.Chatterjee.2011}, which is then quickly connected to a strange-metal regime with a crossover temperature much smaller than a usual Fermi degenerate temperature.

In a BCS framework, the pairing force between the quasiparticles is also important. That is, a $d$-wave pairing of the quasiparticles should be caused by some independent bosonic excitations for a non-phonon SC mechanism \cite{Hirsch.Scalapino.1986, Scalapino.Scalapino.2012}. Namely, distinct \emph{gapped} elementary excitations or fluctuations beyond the gapless quasiparticles would be expected to exist in the pseudogap phase. For instance, the spin resonance mode observed in the cuprate \cite{PhysRevLett.75.316, Dai1999, Keimer.Fong.1999, Keimer.He.20006wc, Keimer.He.2002,Keimer.Capogna.2007, Bourges.Fauqu.2007} has been conjectured as a candidate of the pairing glue in the literature. Nevertheless, the origin of the Fermi arc and the bosonic excitations in the pseudogap phase by a \emph{unified} microscopic description remains a challenging issue. The complexity of the pseudogap physics as competing or/and intertwined orders has been also extensively discussed as the phenomenology in a generalized Landau paradigm \cite{Tranquada.Fradkin.2015doan, Zaanen.Keimer.2015}.

\begin{figure}[tb]
	\centering
	\includegraphics[width=\linewidth]{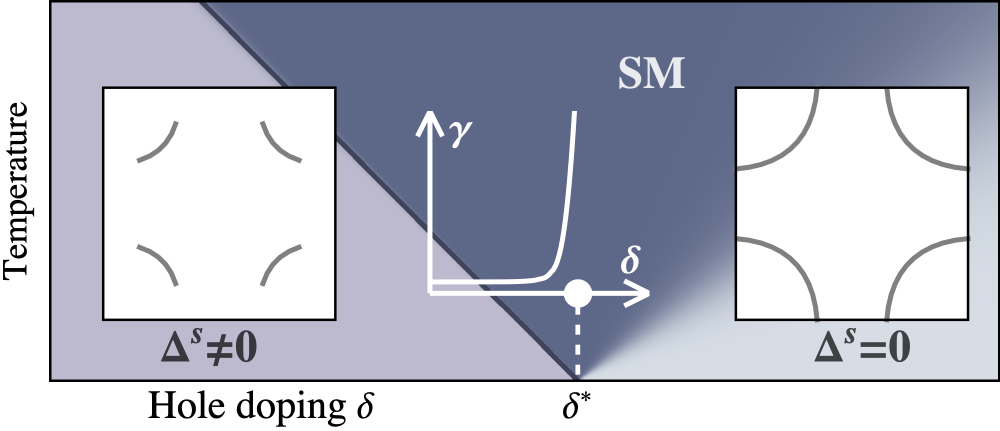}
  \caption{The phase diagram of a pseudogap phase and an overdoped normal state characterized by an RVB order parameter $\Delta^s\neq 0$ at $\delta<\delta^*$ and $\Delta^s= 0$ at $\delta>\delta^*$, respectively, with removing superconducting instability at low-$T$ (Ref. \onlinecite{Weng.Ma.2014}). As to be shown in this work, a quasiparticle excitation at the RPA level exhibits a Fermi arc structure (the left inset), which recovers a full Fermi surface at $\Delta^s= 0$ (the right inset). The Fermi surface crossover occurs at the critical point $\delta^*$, which is marked by a divergent specific heat $\gamma\equiv C_v/T$ (the middle inset) of the background spins as $\Delta^s\rightarrow 0$. SM denotes a high temperature regime known as the strange-metal phase \cite{Weng.Ma.2014}.}
	\label{fig:phase_diagram}
\end{figure}

Alternatively, the pseudogap phase may be regarded as a ``normal state'' of a doped Mott insulator, in which the resonating-valence-bond (RVB) pairing \cite{Anderson.Anderson.1987, Anderson.Baskaran.1987, Shiba.Zhang.1988, Zhang.Anderson.2004,Wen.Lee.2006z4} of the localized electrons, due to the strong on-site Coulomb repulsion, plays a crucial role. From such a perspective, the spin-singlet RVB pairing of the background spins may give rise to not only the much needed non-phonon pairing strength for the high-$T_c$ superconductivity \cite{Anderson.Anderson.1987}, but also a diversity of the pseudogap phenomenon \cite{Wen.Lee.2006z4}. To unify the pseudogap, SC, and antiferromagnetic (AFM) phases in the doped Mott insulator, a more precise two-component RVB description has been proposed
\cite{Weng.Weng.2011, Weng.Ma.2014} based on the $t$-$J$ model with revealing a hidden singular sign structure \cite{Ting.Weng.1997, Overbosch.Zaanen.2011}. In the AFM phase, a bosonic RVB works well \cite{Anderson.Liang.1988}, while in the pseudogap/SC regime at finite doping, the doped holes will introduce a secondary \emph{fermionic} RVB pairing between the dopants. At finite doping, the ground states of the pseudogap and SC phases are essentially the same at temperature $T=0$, while in the former the SC phase coherence is disordered at $T>0$ by the thermal topological defects, but in the latter the SC phase coherence is maintained up to $T\leq T_c$ until a topological transition \cite{Weng.Mei.20107w} to the former. Furthermore, the primary bosonic RVB order parameter $\Delta^s$ vanishes at doping $\delta>\delta^*$, which defines an overdoped regime that may be smoothly connected to a strange-metal phase at high-$T$ as a crossover.  The mean-field phase diagram without a topological SC transition at low $T$ is illustrated in Fig. \ref{fig:phase_diagram} for such a two-component RVB state \cite{Weng.Ma.2014}.

The fate of the quasiparticle excitation is an important issue of the doped Mott insulator. In the one-component RVB theory, the low-lying Landau's quasiparticle, carrying both charge and spin, has been explored as the bound state of the fractionalized particles, i.e., holon and spinon, via effective attractive interactions \cite{Wen.Lee.1998, Ng.Ng.2005}. On the other hand, a phenomenological scheme has been recently employed in the study of the single-particle Green's function for the two-component RVB state, where a Fermi arc structure in the pseudogap phase and a two-gap feature in the SC phase have been identified \cite{Weng.Zhang.2020} in good comparison with the experiments \cite{PhysRevB.98.140507,Ding.Zhong.20197md}. Here the quasiparticle is no longer \emph{elementary} like in a Fermi liquid but rather emergent as a \emph{collective mode} \cite{PhysRevB.61.12328} in a doped Mott insulator. But the following basic questions still need to be carefully examined and answered: What is the emergent nature of the quasiparticle excitation in a fractionalized system and if the Luttinger theorem \cite{Luttinger.Luttinger.1960xws, Ward.Luttinger.1960} is still valid or violated, and especially how can it evolve into a conventional Landau quasiparticle with a full Fermi surface in the overdoped regime?

In this paper, we shall further inspect the gapless quasiparticle excitation and its physical implications based on the \emph{parent} two-component RVB mean-field description in the $t$-$J$ model \cite{Weng.Weng.2011, Weng.Ma.2014}. The single-particle Green's function is systematically formulated by a generalized random-phase-approximation (RPA) scheme, in which a gapless quasiparticle mode emerges along four Fermi arcs in the Brillouin zone in the low-$T$ pseudogap state (without SC phase coherence). With $\Delta^s$ vanishing beyond the critical doping $\delta^*$, a full Fermi surface is recovered. The crossover from the Fermi arc to a full Fermi surface for the gapless quasiparticle excitation  (cf. the insets of Fig. \ref{fig:phase_diagram}) reflects the underlying transition of the background spins from an RVB state to a semiclassical paramagnetic state. The divergent specific heat contributed by the background spins indicated in the middle inset of Fig. \ref{fig:phase_diagram} is a further manifestation of such an RVB transition at $\delta= \delta^*$.

An important conclusion of the analytic analysis of the single-particle Green's function is that the gapless quasiparticle excitation still satisfies the Luttinger volume with a large Fermi surface contour in the momentum space in both pseudogap and overdoped paramagnetic phases. The ``Fermi arc'' feature manifested at $\delta<\delta^*$ is actually due to a strong suppression of the quasiparticle spectral weight by decaying into the fractionalized ``spinon'' in the
antinodal region. In contrast, the low-lying quasiparticle is well protected near the nodal region without a further decay such that a coherent ``Fermi arc'' is preserved as the residual piece of the original large Fermi surface in this region. 

Here the Fermi-arc structure is stable within a region covered by a gapped Fermi pocket in the mean-field state of the fermionic spinon component \cite{Weng.Ma.2014}, which is centered at momenta $(\pm \pi/2, \pm \pi/2)$. Namely the emergent coherent quasiparticle at the RPA level is effectively protected by the underlying mean-field state. We show that the latter fractionalization may be detected experimentally as follows. In a strong magnetic field, the pairing of the fermionic spinons may be diminished inside the magnetic vortex cores \cite{Weng.Ma.2014} such that the Fermi arc of the gapless quasiparticles can get hybridized with the Fermi pocket to give rise to a quantum oscillation, which is in agreement with the experiment \cite{Taillefer.Doiron-Leyraud.2007, Greven.Barii.2013}. 

Another unique structure in this fractionalization mechanism is the appearance of sharply enhanced ``hotspots'' as the endpoints of the ``Fermi arc''. One finds that the scatterings between the hotspots can further lead to a static quasiparticle interference (QPI) pattern, which resembles a weak charge-density-wave (CDW) even though there is no true spontaneous translational symmetry breaking.

Furthermore, besides the fractionalization of a quasiparticle into the fermionic spinon, the bosonic spinons can be also thermally excited by breaking up the bosonic RVB order in the pseudogap phase, which carry vortices in the SC order parameter. It has been previously shown \cite{Weng.Zhang.2020}  that the Fermi arc will be replaced by a $d$-wave-like gap with the Landau quasiparticle turning into a Bogoliubov quasiparticle once the SC phase coherence is realized by logarithmically binding these bosonic spinons into $S=1$ resonance-like modes at $T\leq T_c$. In particular, as $\delta\rightarrow \delta^*$, both the energy scale and the bandwidth of the bosonic spinons get diminished to result in a divergent specific heat as indicated in Fig. \ref{fig:phase_diagram}.

On the other hand, the decay of a quasiparticle excitation can be effectively stopped  at $\delta>\delta^*$ due to the total disordering of the background spins with $\Delta^s=0$, which are thermalized in a semiclassical Curie-Weiss-like state.  Then a large Fermi surface can be explicitly manifested in the spectral function. In this sense, the phase at $\delta>\delta^*$ resembles a Fermi liquid state as the normal state at low-$T$. However, the overall spectral weight for the quasiparticle is much reduced, because the background spin degrees of freedom and incoherent spinless ``holons'', although they are effectively decoupled from the quasiparticles at low temperature, are still present and even dominate at higher temperatures in the so-called strange-metal region.

%which is called the \cite{Weng.Sheng.1996, Ting.Weng.1997, Overbosch.Zaanen.2011}

The rest of the paper is organized as follows. In \secref{sec2}, we
briefly outline the nontrivial phase-string sign structure of the $t$-$J$ model and the corresponding fractionalization scheme of the two-component RVB state.
Then, a generalized RPA diagrammatic Dyson equation for the single-particle Green's function is presented. In \secref{sec3}, we systematically explore the single-particle Green's function in both the pseudogap phase at $\delta<\delta^*$ (\secref{sec3A}) and overdoped phase at $\delta>\delta^*$ at $T=0$ (\secref{sec3B}). In particular, the hotspots as the enhanced density of states are identified near the end of the Fermi arcs in the underdoped regime (\secref{sec3A3}), which lead to a novel CDW-like QPI pattern (\secref{secCDW}). Furthermore, with the disappearance of the RVB order at $\delta^*$, a divergent specific heat is found in the spin background as shown in \secref{sec3B1}. In the discussion section (\secref{sec6}), different approaches on the origin of the Fermi arc are discussed. In particular, it is pointed out that the Luttinger volume is always satisfied on both sides of the critical point $\delta^*$ in the present work, which is fundamentally distinct from the other schemes. However, we also show that a Fermi pocket structure violating the Luttinger volume does appear in a special limit of our theory, when the secondary RVB order vanishes first in a strong magnetic field.
Finally, the conclusion and perspectives are given in \secref{sec7}.

\section{Effective theoretical framework} \label{sec2}
\subsection{Phase-string effect and two-component RVB mean-field state in the $t$-$J$ model}
The $t$-$J$ model on a 2D square lattice is given by $H_{t\text{-}J}=H_{t}+H_{J}$, where
\begin{eqnarray}\label{Ht}
  H_{t}&=&-t \sum_{\langle i j\rangle \sigma} c_{i \sigma}^{\dagger} c_{j \sigma}+\text {h.c.}, \\
	H_{J}&=& {J} \sum_{\langle i j\rangle}\left(\boldsymbol{S}_{i} \cdot \boldsymbol{S}_{j}-\frac{1}{4} n_{i} n_{j}\right), \label{HJ}
\end{eqnarray}
in which $\boldsymbol{S}_i$ and $n_{i}=\sum_{\sigma} c_{i \sigma}^{\dagger} c_{i \sigma}$ are the local $SU(2)$ spin operator and electron number operator, respectively. The Hilbert space is restricted by the no double occupancy constraint at any site $i$
\begin{equation}\label{ori_cons}
	n_{i} \leqslant 1.
\end{equation}

The $t$-$J$ model has been widely considered as a minimal model for the high-$T_c$ cuprate. Due to the no double occupancy constraint Eq. (\ref{ori_cons}), the conventional Fermi statistical sign structure will reduce to the phase-string sign structure \cite{Weng.Sheng.1996,Zaanen.Wu.2008} in the $t$-$J$ model. In particular, it is statistical sign-free at half-filling, where the ground state as governed by the Heisenberg model $H_J$ in Eq. (\ref{HJ}) only possesses a trivial Marshall sign \cite{Marshall.Marshall} with spins forming an antiferromagnetic (AFM) long-range order in the thermodynamic limit. The real challenge is at a finite doping, where SC and pseudogap phases are expected to appear in the underdoped regime, where the AFM long-range order disappears.

Upon doping, the singular phase-string effect will prevent a coherent propagation of the bare doped holes\cite{Weng.Sheng.1996, Ting.Weng.1997}.  The starting point of the present work will be based on a duality transformation to explicitly incorporate the phase-string sign structure, which leads to a peculiar fractionalization scheme \cite{Weng.Weng.2011} distinct from the usual slave-particle approaches \cite{Nagaosa.Lee.1992, Wen.Lee.1998, Wen.Lee.2006z4}.
The ground state can be generally written in the following form \cite{Weng.Weng.2011,Weng.Ma.2014}
\begin{equation}\label{parent}
	|\Psi_G\rangle=e^{i\hat{\Theta}}\left [\hat{\mathcal{P}}\left|\Phi_{h}\right\rangle \otimes\left|\Phi_{a}\right\rangle \otimes \left |\Phi_b\right \rangle\right]
\end{equation}
where $e^{i\hat{\Theta}}$ is a unitary transformation due to the statistical or phase-string sign structure, which leads to a duality world in which the electrons are fractionalized into bosonic $h$-holons described by $|\Phi_{h}\rangle$ and two-component spinon state characterized by $|\Phi_{a}\rangle \otimes |\Phi_b\rangle$. In the latter, $|\Phi_b\rangle$ is an RVB state composed of the bosonic $b$-spinons, which is further supplemented by \emph{fermionic} backflow $a$-spinons introduced by holes, which are in a BCS-like pairing state $|\Phi_{a}\rangle$  (see below). At half-filling, $|\Psi_G\rangle\rightarrow \hat{\mathcal{P}} |\Phi_b \rangle$, which naturally reduces to the AFM ground state of the Heisenberg model with $e^{i\hat{\Theta}}\rightarrow 1$. At finite doping, the Bose-condensation of the holons in $|\Phi_{h}\rangle$ together with the hidden off-diagonal-long-range-order (ODLRO) in the aforementioned two-component RVB state will constitute the important partial rigidity (fractionalization) to characterize the so-called \emph{lower} pseudogap phase (LPP)  \cite{Weng.Ma.2014} by Eq. (\ref{parent}). Figure \ref{fig:phase_diagram} illustrates the general phase diagram obtained in Ref. \onlinecite{Weng.Ma.2014}.

Here the duality transformation in Eq. (\ref{parent}) is critical to accommodate the phase-string sign structure, with
\begin{eqnarray}
         \hat{\Theta} \equiv -\sum_i n^h_i \hat{\Omega}_{i}
\end{eqnarray}
which may be interpreted as each doped hole introduces a nonlocal phase-shift $\hat{\Omega}_{i} $ in the $b$-spinon background  $|\Phi_{b}\rangle$. Here $n_i^h$ denotes the holon number at site $i$, and
\begin{eqnarray}
        	\hat{\Omega}_{i} &\equiv& \frac{1}{2}\left(\Phi_{i}^{s}-\Phi_{i}^{0}\right),\\
	\Phi_{i}^{s}&=&\sum_{l \neq i} \theta_{i}(l)\left(n_{l \uparrow}^{b}-n_{l \downarrow}^{b}\right),\label{vortexb}\\
	\Phi_{i}^{0}&=&\sum_{l \neq i} \theta_{i}(l),
\end{eqnarray}
with $n_{l \sigma}^{b}$ denoting the number operator of the $b$-spinon at site $l$, and $\theta_{i}(l) \equiv \pm \operatorname{Im} \ln \left(z_{i}-z_{l}\right)$ ($z_i$ is the complex coordinate of site $i$). Physically how the phase-shift $\hat{\Omega}_{i} $ emerges due to the phase-string can be transparently seen in the studies of the single-hole and two-hole doped ground states \cite{Weng.Chen.2018, Weng.Chen.2019wx, Weng.Zhao.2022}. %\cite{Weng.Zhu.2014,  Weng.Zheng.2018,Weng.Zhu.2018ss, Weng.Chen.2018, Weng.Chen.2019wx, Weng.Zhao.2022}.

After such a duality transformation, the resulting three subsystems in Eq. (\ref{parent}) are quite conventional to describe the LPP as given by \cite{Weng.Weng.2011,Weng.Ma.2014}:
\begin{equation}
|\Phi _{h}\rangle \equiv \sum_{\{l_{h}\}}\varphi
_{h}(l_{1},l_{2},...)h_{l_{1}}^{\dagger }h_{l_{2}}^{\dagger }...|0\rangle
_{h}\ ,  \label{bgs}
\end{equation}%
and
\begin{equation}
|\Phi _{a}\rangle \equiv \exp \left( -\sum_{ij}{g}_{ij}a_{i\uparrow
}^{\dagger }a_{j\downarrow }^{\dagger }\right) |0\rangle _{a}~,  \label{phia-0}
\end{equation}%
and
\begin{equation}
|\Phi _{b}\rangle \equiv \exp \left( \sum_{ij} W_{ij}b_{i\uparrow }^{\dagger
}b_{j\downarrow }^{\dagger }\right) |0\rangle _{b}~.  \label{phirvb}
\end{equation}
The corresponding effective Hamiltonian after the duality transformation is given in \appref{app_mean-field}, in which the fractionalized states given above are determined at a generalized mean-field level \cite{Weng.Ma.2014}.
Here the two-component RVB state is characterized by a BCS-like pairing amplitude ${g}_{ij}$ with an $s$-wave-like order parameter $\Delta^a$ and a bosonic RVB pairing amplitude $W_{ij}$ characterized by an order parameter $\Delta^s$ which reduces to the conventional Schwinger-boson mean-field order parameter at half-filling limit without holes. In such an RVB regime at low doping, the holon wavefunction $\varphi _{h}= \text{constant} $ indicates a Bose-condensed holon state $|\Phi _{h}\rangle $.

\begin{figure}[t]
	\centering
	\includegraphics[width=\linewidth]{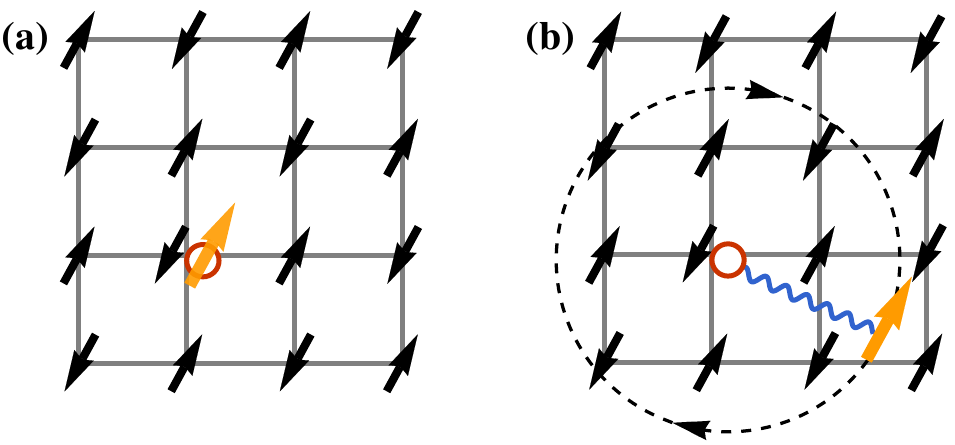}
	\caption{Schematic illustration of the quasiparticle fractionalization in \eqnref{Eq:1} and \eqnref{ctilde}. (a) A bare hole created by $\hat{c}_i$: A black arrow represents a background $b$-spinon and an orange arrow denotes the backflow $a$-spinon associated with the hole, which makes the total spin zero at the hole site; (b) A  ``twisted'' hole of $\tilde{c}_i$ is a mobile composite with the spin and charge partners forming an internal relative motion, as indicated by the blue wavy line and a dashed circle with arrows, which facilitates the coherent motion of the hole in a singlet RVB background of the $b$-spinons.}
	 	\label{fig_twist}
\end{figure}

The projection operator $\hat{\mathcal{P}}$ in Eq. (\ref{parent}) enforces the original no double occupancy constraint such that
\begin{eqnarray}\label{projection}
	\sum_\sigma n_{i \sigma}^{b}=1, ~~~~ n_{i\bar\sigma}^{a}=n_{i}^{h} n_{i \sigma}^{b}.
\end{eqnarray}
Here $n_{i\bar\sigma}^{a}$ is the number of the backflow $a$-spinons with spin $\bar\sigma=-\sigma$ at site $i$, and as defined before, $n_i^h$ is the holon number and $n_{i \sigma}^{b}$ is the number operator of the $b$-spinons.
Namely under the projection, in the two-component spinon description, the $b$-spinons will always remain singly-occupied per lattice site, whereas the backflow $a$-spinon compensates the total spin at each hole site, i.e., $n_{i}^{h}\boldsymbol{S}_{i}^{b} +\boldsymbol{S}_{i}^{a}=0$ for the $S^z$-component - which is always chosen as the quantization axis in the present formulation - while the ${S}^{x,y}$ components as the $U(1)$ phase can be always absorbed by the $h$-holons \cite{Weng.Weng.2011,Weng.Ma.2014}. As illustrated in \figref{fig_twist}(a) where black arrows represent the $b$-spinons, which at the hole sites are compensated by the $a$-spinons (orange arrow). Such constraints in Eq. (\ref{projection}) are enforced at the mean-field level by introducing the Lagrangian multipliers \cite{Weng.Weng.2011,Weng.Ma.2014} (cf. \appref{app_mean-field}).

It is important to note that the usual $U(1)$ gauge fluctuations associated with the constraints in Eq. (\ref{projection}), which is in the transverse current channels beyond the projection $\hat{\mathcal{P}}$, are all suppressed (Higgsed) in the LPP due to the ODLROs in Eqs. (\ref{bgs}), (\ref{phia-0}) and (\ref{phirvb}), i.e., the holon condensation and two-component RVB orders. In other words, in the present LPP state, the fractionalization is self-consistently protected by the hidden ODLROs in the subsystems. On the other hand, these fractional elementary particles are still weakly coupled (between the $h$-holons and $b$-spinons) via the mutual Chern-Simons gauge fields as the topological gauge fields in the dual world due to the irreparable phase-string \cite{Weng.Kou.2003yuc, Weng.Qi.2007}.

To summarize, in the above we have outlined the previously developed phase-string mean-field formulation of the $t$-$J$ model with the detailed formalism given in Appendex \ref{app_mean-field} and \ref{sec_App_selfconsis}, which will be the starting point for the present study of the single-particle excitation. In order to have a better understanding of the phase-string formalism, in the following, we make a critical comparison of the mean-field state in Eq. (\ref{parent}) with the conventional slave-boson mean-field state \cite{Wen.Lee.2006z4} given as follows: 
\begin{equation}
	|\Psi_{\mathrm {RVB}} \rangle=\hat{\mathcal{P}}\left [\left|\Phi_{h}\right\rangle \otimes \left |\Phi_{\mathrm {f-RVB}}\right \rangle \right], \nonumber
\end{equation}   
where $\left|\Phi_{\mathrm {f-RVB}}\right \rangle$ denotes a BCS-like fermionic spinon RVB state, which replaces the two-component RVB state given in Eq. (\ref{parent}) with the Gutzwiller projection operator $\hat{\mathcal{P}}$ enforcing the no-double-occupancy constraint. The most prominent distinction is that the nonlocal mutual duality transformation $e^{i\hat{\Theta}}$ in Eq. (\ref{parent}) is absent in the above slave-boson mean-field state. Here, if the holons are Bose-condensed in $\left|\Phi_{h}\right\rangle$,  $|\Psi_{\mathrm {RVB}} \rangle$ is always superconducting. By contrast, even with the Bose-condensation of the holons and the RVB pairings of the two-component spinons, Eq. (\ref{parent}) is intrinsically non-superconducting at $T=0^+$ as the thermally excited $b$-spinons in $\left|\Phi_{b}\right\rangle$ will always disorder the phase coherence of $|\Psi_G\rangle$ via the phase factor $e^{i\hat{\Theta}}$. The true superconducting phase coherence is only realized at a lower temperature via a further topological transition, which is not the main focus of this work (see the following subsection). 

Therefore, with explicitly incorporating the phase-string sign structure of the $t$-$J$ model via the duality transformation, a new non-superconducting state, i.e., LPP, will emerge to replace the superconducting phase in the slave-boson mean-field state to describe the pseudogap phase at low-temperature. In particular, in this phase-string formalism, the bosonic RVB state $\left|\Phi_{b}\right\rangle$ can accurately recover the antiferromagnetic ground state at half-filling where the fermionic $a$-spinons disappear. At finite doping, in contrast to the full fermion RVB state $\left|\Phi_{\mathrm {f-RVB}}\right \rangle$, however, one may naturally ask about the fate of a quasiparticle excitation, which is created by the electronic $c$-operator in the LPP. In the following, we shall discuss the issue of the single-particle (hole) excitation, which will go beyond the the fractionalized mean-field state of Eq. (\ref{parent}).

\subsection{ Quasiparticle as an emergent ``collective'' excitation in the LPP}

Due to the phase-string sign structure, the strongly correlated electrons in the doped Mott insulator are fractionalized into more elementary constituents of the bosonic $h$-holons, bosonic $b$-spinons, and fermionic backflow $a$-spinons. They are coherent and form ODLROs in the LPP, i.e., the holon condensation and two-component RVB pairings. In turn, the fractionalization is further strengthened by the ODLROs as the usual internal $U(1)$ fluctuations associated with the decomposition get Higgsed (not the external $U(1)$ associated with the electromagnetic field, see below). Therefore, self-consistently the phase-string leads to a peculiar electron fractionalization with the elementary particles of $h^{\dagger}$, $b^{\dagger}$ and $a^{\dagger}$, which are further protected by the hidden ODLROs.

Then the Landau quasiparticle, which is the most important elementary excitation in a Landau FL state, is not present as a stable object at the mean-field LPP. As a matter of fact, a \emph{bare} hole (electron) has no trace in the mean-field ground state Eq. (\ref{parent}). What is the fate of the quasiparticle in a fractionalized ground state is generally an important issue  \cite{Wen.Lee.2006z4}.  In the following, we first note that the creation of a bare hole on the ground state will decay (fractionalize) into the elementary particles by \cite{Weng.Weng.2011, Weng.Ma.2014}:
 \begin{eqnarray}\label{Eq:1}
 	\hat{c}_{i\sigma} & = & h_{i}^{\dagger} a_{i\bar \sigma}^{\dagger} e^{i \hat\Omega_{i}}, \notag\\
	& \equiv &  \tilde{c}_{i \sigma} e^{i \hat\Omega_{i}},
 \end{eqnarray}
where
\begin{eqnarray}\label{ctilde}
 	\tilde{c}_{i\sigma} \equiv h_{i}^{\dagger} a_{i\bar \sigma}^{\dagger} . \label{Eq:11}
\end{eqnarray}
(Note that here a trivial sign factor has been absorbed by redefining $a^{\dagger}_{i\bar{\sigma}}$, and the notation $\tilde{c}_{i\sigma} $ is also different from that used in Refs. \onlinecite{Weng.Weng.2011, Weng.Ma.2014}.) In such a novel fractionalization framework, the twisted hole $\tilde{c}_{i \sigma}$ is schematically illustrated in \figref{fig_twist}(b) as a composite, in which the $a$-spinon denotes a spin that is always associated with the holon via the underlying RVB pairing.  Since the holon is condensed in the LPP, thus $\tilde{c}_{i \sigma}$ behaves essentially like the $a$-spinon. The aforementioned ODLROs in the LPP protects the $a$-spinon as a well-defined gapped excitation. In this sense, a bare hole created by the electron $c$-operator in Eq. (\ref{Eq:1}) is composite in the fractionalized LPP. 

According to Eq. (\ref{Eq:1}), the superconducting pairing is characterized by
 \begin{equation}\label{SC}
 	\left\langle \hat{c}_{i\uparrow}\hat{c}_{j\downarrow} \right\rangle \propto  \Delta^0_{ij} \left\langle e^{i \frac 1 2 \left (\Phi^s_{i}+\Phi^s_{j}\right)}\right\rangle ,
 \end{equation}
 where the amplitude $\Delta^0_{ij}\propto \Delta^a_{ij}$ with the holon condensation, while
 \begin{equation}\label{phdisor}
 \left\langle e^{i \frac 1 2 \left (\Phi^s_{i}+\Phi^s_{j}\right)}\right\rangle=0,
 \end{equation}
once \emph{free} $b$-spinons appear in Eq. (\ref{parent}). In other words, even though the superconductivity is naturally realized in the ground state, where the $b$-spinons form a short-range RVB pairing, the phase coherence can be immediately disordered at finite temperature due to the excited $b$-spinons at the mean-field level, which is the true LPP mean-field state studied in Ref. \onlinecite{Weng.Ma.2014} with the preformed pairing $\Delta^0_{ij}\neq 0$. On the other hand, with $\Delta^s=0$ at $\delta>\delta^*$, the phase disordering is expected to occur at $T=0$ in Eq. (\ref{parent}). In both cases, the external $U(1)$ symmetry is not broken, without the SC phase coherence, which will be the main focus of the present work with the quantitative mean-field phase diagram \cite{Weng.Ma.2014} illustrated in Fig. \ref{fig:phase_diagram}.

Previously it has been shown \cite{Weng.Zhang.2020} that a bare hole injected into the LPP does not fractionalize immediately. As shown by the equation-of-motion method, it will propagate coherently for a while before it finally decays according to Eq. (\ref{Eq:1}). In particular, a coherent quasiparticle excitation can even be stabilized within the region that is occupied by the fermionic $a$-spinons, where its decay in terms of Eq. (\ref{Eq:1}) is prevented by the Pauli exclusion principle and is further protected by the $s$-wave gap of the $a$-spinons. A Fermi arc (incomplete Fermi surface) is thus indicated  in the LPP \cite{Weng.Zhang.2020}.

On the other hand, according to the Oshikawa's topological argument \cite{Oshikawa.Oshikawa.2000},  if a gapless single-particle excitation (quasiparticle) with $U(1)$ symmetry and translation
symmetry exists, its Fermi surface should still satisfy the Luttinger volume, i.e., by enclose an area to accommodate
the total number of electrons. In other words, the Landau quasiparticle would be expected to
present as a low-lying excitation even in a strongly correlated system under some very general conditions.

In the following, we shall systematically explore the issue if and how a bare hole created by $\hat{c}$ can maintain its stability as an \emph{emergent} low-lying excitation in the fractionalized state of Eq.(\ref{parent}).  In particular, how its gapless Fermi surface structure evolves from $\delta<\delta^*$ to $\delta>\delta^*$ will be studied and a large Fermi surface satisfying the Luttinger's volume will be reconciled with the Fermi arc phenomenon in the LPP.

\subsection{Dyson equation of the single-particle Green's function}

The single-particle Green's function is defined by
\begin{eqnarray}\label{Eq:2}
G^e\left(\boldsymbol{r}_{i}-\boldsymbol{r}_{j}; \tau\right) \equiv -\left\langle \hat T_{\tau} \hat{c}_{i \sigma}(\tau) \hat{c}_{j \sigma}^{\dagger}(0)\right\rangle,
\end{eqnarray}
where $\hat T_{\tau}$ denotes the imaginary-time ordering operator. At the mean-field level of Eq. (\ref{parent}), it can be further expressed according to Eq.~(\ref{Eq:1}) by
\begin{eqnarray}\label{Eq:3}
G^e_{0}\left(\boldsymbol{r}_{i}-\boldsymbol{r}_{j}; \tau \right) &=& -\left\langle \hat T_{\tau} \tilde{c}_{i \sigma}(\tau) \tilde{c}_{j \sigma}^{\dagger}(0) e^{i\left[\hat\Omega_{i}(\tau)-\hat\Omega_{j}(0)\right]}\right\rangle_0\notag \\
&=&  D_{\tilde{c}}(\boldsymbol{r}_{i}-\boldsymbol{r}_{j};\tau) f(\boldsymbol{r}_{i}-\boldsymbol{r}_{j}; \tau) \notag \\
 &\equiv &  \diag{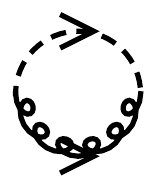}{27pt}
\end{eqnarray}
where
\begin{eqnarray}
D_{\tilde{c}}(\boldsymbol{r}_{i}-\boldsymbol{r}_{j}; \tau) &=&-\left\langle \hat T_{\tau} \tilde{c}_{i \sigma}(\tau) \tilde{c}_{j \sigma}^{\dagger}(0) e^{-i\sum_{i \rightarrow j} \phi_{i_{s} i_{s+1}}^{0}} \right\rangle_{0} \notag \\
& & \ \ \ \times  e^{-i \boldsymbol{k}_{0} \cdot(\boldsymbol{r}_{i}-\boldsymbol{r}_{j})} \notag \\
&\equiv & \diag{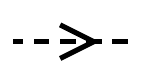}{12pt} \label{Eq:4}\\
f(\boldsymbol{r}_{i}-\boldsymbol{r}_{j}; \tau) &=&\left\langle \hat T_{\tau} e^{\frac{i}{2} \Phi_{i}^{s}(\tau)} e^{-\frac{i}{2} \Phi_{j}^{s}(0)}\right \rangle_{0} \notag \\
&\equiv & \diag{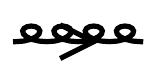}{12pt} \label{f}
\end{eqnarray}
with $\boldsymbol{k}_{0}=(\pm\pi / 2, \pm\pi / 2)$ %or $(-\pi / 2, -\pi / 2)$,
and the phase factor inside the average of \eqnref{Eq:4} coming from $e^{-\frac{i}{2}\left[\Phi_{i}^{0}-\Phi_j^{0}
\right]}$,
which is originated from $e^{i\left[\hat\Omega_{i}-\hat\Omega_{j}\right]}$ to keep $D_{\tilde{c}}$ gauge invariant (cf. \appref{sec:simplify} for the detail).

$G^e_{0}$ here describes the complete fractionalization without the trace of the gapless Landau quasiparticle. On the other hand, based on the equation-of-motion \cite{Weng.Zhang.2020}, a bare hole can propagate coherently based on the original $t$-$J$ model before decaying into the fractionalized elementary particles with a vertex $\lambda $. At the RPA level, the single-particle Green's function may be diagrammatically expressed as
\begin{equation}\label{Dyeq}
  \diag{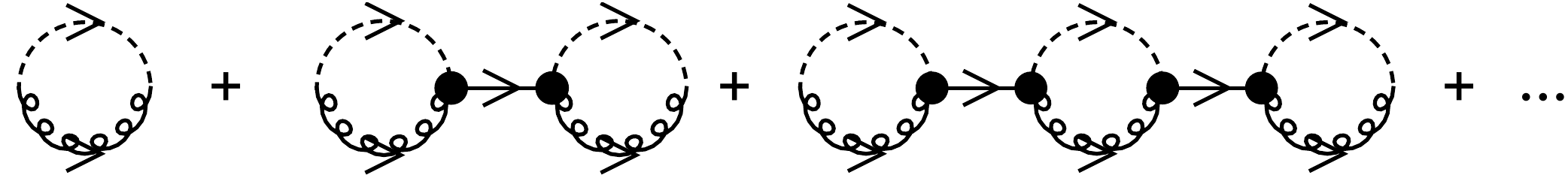}{24pt}
\end{equation}
where the straight lines with arrows in the middles denote the bare quasiparticle propagator:
\begin{equation}
 	\label{Eq:5}
G^c_{0}(\boldsymbol{r}_{i}-\boldsymbol{r}_{j}; \tau) \equiv \diag{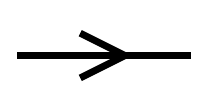}{12pt}
\end{equation}
with a large Fermi surface satisfying the Luttinger volume as predicted by a conventional band theory. Namely, the general Dyson equation is given by
\begin{eqnarray}
    &\;&G^e(\boldsymbol{r}_{i}-\boldsymbol{r}_{j}; \tau)=G^e_0(\boldsymbol{r}_{i}-\boldsymbol{r}_{j}; \tau)  \notag\\
&\;&\;\;\;+\lambda^{2} \iint d \tau^{\prime} d \tau^{\prime \prime} \sum_{\boldsymbol{r}_{i^{\prime}},\boldsymbol{r}_{i^{\prime\prime}}} G^e_0\left(\boldsymbol{r}_i-\boldsymbol{r}_{i^{\prime}};\tau-\tau^\prime \right) \notag\\
  &\;&\;\;\;\times  G^c_0\left(\boldsymbol{r}_{i^{\prime}}- \boldsymbol{r}_{i^{\prime \prime}};\tau^\prime-\tau^{\prime\prime}\right) G^e_0\left(\boldsymbol{r}_{i^{\prime \prime}}- \boldsymbol{r}_j; \tau^{\prime\prime}\right)+\ldots \label{Eq:6}
\end{eqnarray}
where the vertex $\diag{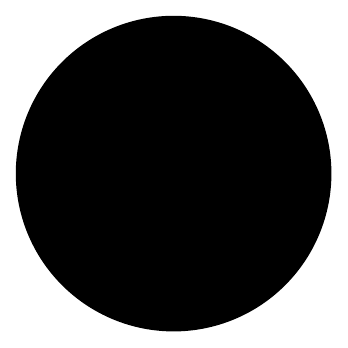}{5pt}=\lambda$ in \eqnref{Dyeq} is approximately assumed to be a constant, which has been estimated \cite{Weng.Zhang.2020} $\sim J\delta$. In the momentum and frequency space, the above single-particle Green's function at the generalized RPA level can be further written in a compact form
\begin{equation}
 	\label{Eq:7}
G^e(\boldsymbol k;\omega) = \frac 1{{G^e_0}^{-1}(\boldsymbol k; \omega)- \lambda^{2} G^c_0(\boldsymbol k; \omega)}.
\end{equation}

Note that the Dyson equation in \eqnref{Dyeq} is slightly different from the RPA diagram in the earlier work \cite{Weng.Zhang.2020} for the propagator of $\tilde{c}$, which is in a matrix form under the Nambu representation even in the LPP phase, since the backflow spinons of $\tilde{c}$ are paired. In contrast, under the present RPA formalism for the gauge-invariant $\hat{c}$, the single-particle Green's function in the LPP resumes a simpler scalar form in Eq. (\ref{Eq:7}), which is also easier for an analytic analysis. The results of the two approaches should become equivalent at higher energies. The present formulation is also similarly connected to the Nambu matrix representation in the SC phase (cf. \appref{sec_App_SC} for more discussions of the SC phase).

%Since some high-energy results (e.g. ``kink'' behavior\cite{Shen.Bogdanov.2000, Shen.Lanzara.2001, Ding.Zhong.20197md} in the LPP, ``two-gap structure''\cite{Kotliar.Civelli.2008, Shen.Hashimoto.2014} in SC phase) are not sensitive to this correction, here we focus on near-zero-energy physical behavior, while more consequences at higher energy can still refer to Ref. \onlinecite{Weng.Zhang.2020}.}

%Therefore, the single-particle excitation is characterized by the phenomenological description of Eq. (\ref{Eq:7}) as a collective mode emerging out of the elementary particles of $h^{\dagger}$, $b^{\dagger}$ and $a^{\dagger}$ in the mean-field fractionalized ground state in Eq. (\ref{parent}).

\begin{figure*}[tb]
 \includegraphics[width=\linewidth]{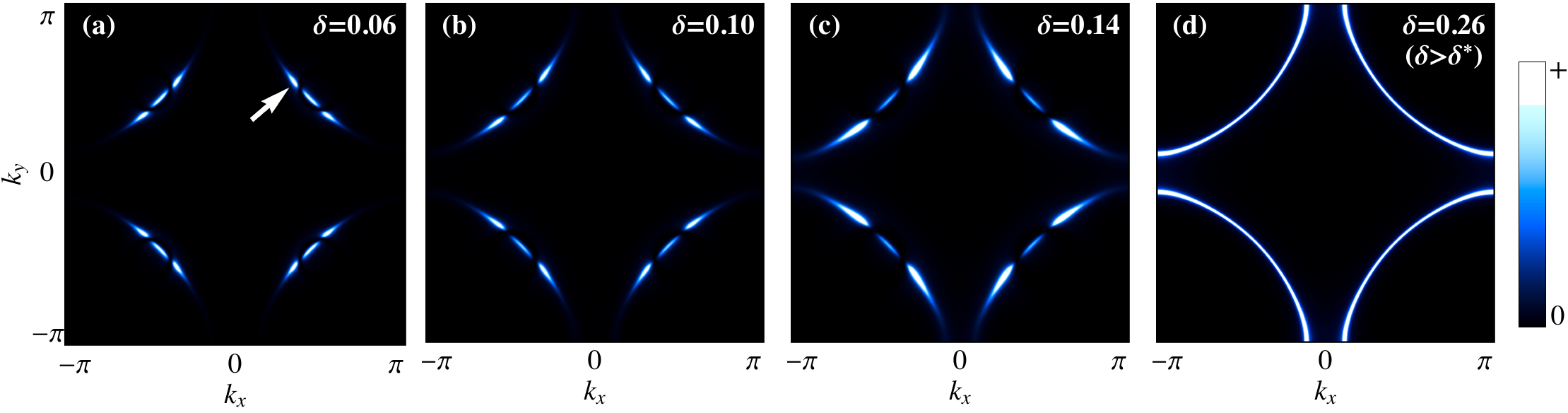}
 \caption{The quasiparticle spectral function $A(\boldsymbol k,\omega)$ at the Fermi level $\omega=0$.
 (a-c) A systematic evolution of the Fermi arcs calculated by \eqnref{GLPG} at various dopings in the LPP ($\delta<\delta^*$). The white arrow in (a) denotes a ``hotspot'' with enhanced spectral weight; (d) A full Fermi surface at $\delta>\delta^*$ based on \eqnref{GFL}. Here $\delta^*\simeq 0.26$ is determined by the set of parameters used in the mean-field self-consistent equations of Ref. \onlinecite{Weng.Ma.2014} .
 }
 \label{fig:arcevo}
\end{figure*}

%In the following text, we will discuss the different behaviors of quasiparticles $c_i$ by simplifying this Dyson equation in distinct phases.After introducing this phenomenological vertex, now we still treat \eqnref{Eq:3} as the leading term and consider the RPA level correction for $G_c$ involving infinite decay and recombination processes with amplitude $\lambda$, finally give the renormalized single-particle Green's function:

%. And it is these residual interactions have probability to bound the fractionalized particles together and form Landau quasi-particles. According to the Oshikawa's topological argument\cite{Oshikawa.Oshikawa.2000} for Luttinger's theorem, if such coherent gapless quasi-particle excitations exist in a system with $U(1)$ symmetry as well as translational symmetry, the Fermi surface will form a closed curve with a total area equal to the total number of electrons. Therefore, we apply the phenomenological decay process $\diag{vertex}{19pt}$ with

%%%%%%%%%%%%%%%%%%%%%%%%%%%%%%%%%%%%%%%%%%%%%%%%%%%%%%%%%%%%%%%%%%%%%%%%%%%%%%%%%%%%%%%%%%%%%%
\begin{figure}[b]
	\centering
	\includegraphics[width=\linewidth]{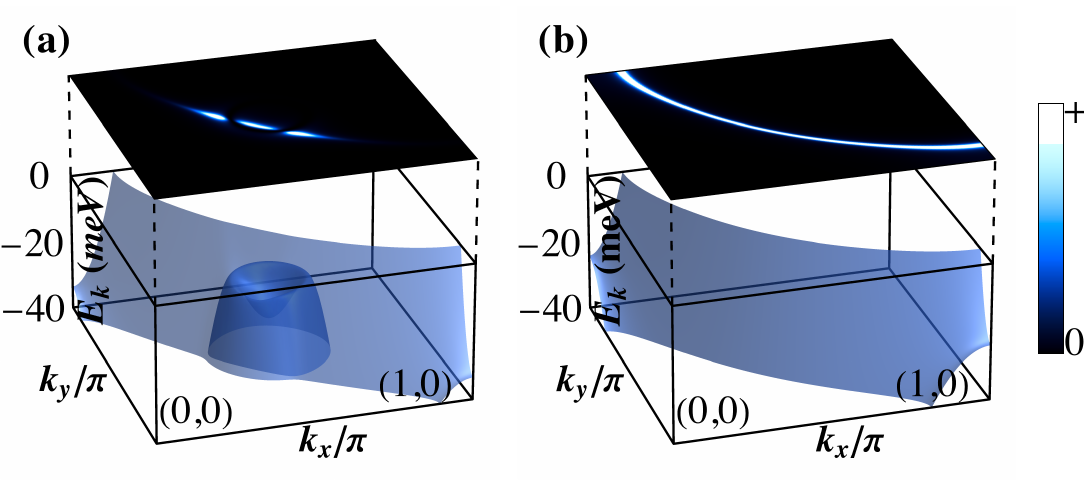}
	\caption{
		(a) Lower part: The poles of \eqnref{GLPG} given in \eqnref{El} in the first quadrant of the Brillouin zone at $\delta=0.06<\delta^*$; Top: Corresponding $A(\boldsymbol k,\omega=0)$; (b) $\delta>\delta^*$.
	}
	\label{singularity}
\end{figure}

%\begin{figure*}[tb]
%	\centering
%		\includegraphics[width=\linewidth]{fig_weight.pdf}
%	\caption{
%		(a)Illustration of the relative position between zero-energy singularity of quasiparticle (red line) and the $a$-spinon pockets(blue circle) without BCS-type pairing $\Delta^a$ centered at $(\pm\pi/2, \pm\pi/2)$, with their cross points labeled by the dark blue point at $\boldsymbol k^*$. (b) The renormalized spectral weight obtained by \eqnref{weight} along the zero-energy singularity contour with angle $\theta$ as marked in (a) at $\delta=0.06$(blue dot) and $\delta=0.14$(red square). The position where spectral weight is suddenly suppressed is just $\boldsymbol k^*$ as labeled in (a).
%	}
%	\label{fig_weight}
%\end{figure*}

%\section{Lower pseudogap phase at $\delta<\delta^*$}\label{sec3}
\section{Quasiparticle excitations with emergent Fermi surfaces}\label{sec3}

In Section \ref{sec2}, the parent phase of the $t$-$J$ model has been formulated in terms of the elementary particles of holon $h^{\dagger}$ and two-component spinons, $b^{\dagger}$ and $a^{\dagger}$, based on the phase-string sign structure. The corresponding mean-field state in Eq. (\ref{parent}), as characterized by their ODLROs, gives rise to a phase diagram illustrated in Fig. \ref{fig:phase_diagram}. A Landau quasiparticle excitation is no longer an intrinsic excitation at such a fractionalized mean-field level, but as a composite object defined in Eq. (\ref{Eq:1}), it may reemerge as a ``collective mode'' at the RPA level. In the following, we shall examine such a mode based on the generalized RPA scheme presented above in Eq. (\ref{Eq:7}).

\subsection{Fermi arc at $\delta<\delta^*$}\label{sec3A}

The LPP at $\delta<\delta^*$ is characterized by the two-component RVB order parameters, $\Delta^s$ and $\Delta^a$, and the holon condensation $\langle h^{\dagger}_{i}\rangle \neq 0$. But it is short of a true ODLRO of SC or AFLRO as we focus on the case that the phase coherence is disordered in Eq. (\ref{phdisor}) while the AFM correlation length is finite \cite{Weng.Ma.2014}.

The corresponding mean-field (fractionalized) Green's function [Eq.(\ref{Eq:3})] in the momentum and frequency space can be expressed by
\begin{equation}\label{GPG}
G^e_0(\boldsymbol k; \omega)\simeq -\mathcal{D}_a(-\boldsymbol k - \boldsymbol k_{0};-\omega){F_0},
\end{equation}
by choosing the gauge of $\phi^0_{ij}$ of the $\pi$-flux such that $e^{-i \sum_{i \rightarrow j} \phi_{i_{s} i_{s+1}}^{0}}=(-1)^{i_x-j_x}$ or $(-1)^{i_y-j_y}$ in Eq. (\ref{Eq:4}), which can be simply absorbed into $e^{-i \boldsymbol{k}_{0} \cdot\left(\boldsymbol{r}_{i}-\boldsymbol{r}_{j}\right)}$ such that $D_{\tilde c}(\boldsymbol k;\omega)\rightarrow -\mathcal{D}_{a}(-\boldsymbol k - \boldsymbol k_{0};-\omega)$ noting the holon condensation. Here $\mathcal{D}_{a}$ is the Green's function of the $a$-spinons which are in BCS-like state [cf. Eq. (\ref{phia-0})] as given by
\begin{equation}\label{GaPG}
  \mathcal{D}_a(\boldsymbol k; \omega)=\frac{1}{\omega-\xi_{ k}^a-\Delta_{ k}^{2} /\left(\omega+\xi_{k}^a\right)}
\end{equation}
where $\xi_{k}^a=-2(t_{a}+\gamma \chi^{a}) \sqrt{\cos ^{2} k_{x}+\cos ^{2} k_{y}}+\lambda_{a}$ and $\Delta_{k}=2  \Delta^{a} \sqrt{\cos ^{2} k_{x}+\cos ^{2}k_{y}}$ is the dispersion for the $a$-spinons and the $s$-wave BCS pairing order parameter [cf. \appref{app_mean-field}].

Furthermore, due to the short-range singlet pairing of the $b$-spinons in the LPP state, $f(\boldsymbol{r}_{i}-\boldsymbol{r}_{j}, \tau_{i}-\tau_{j})$ in the large distance may be approximately reduced to a constant $F_0$ as given in Eq. (\ref{GPG}) at $T=0$. However, as it is pointed out previously, the LPP state in Eq. (\ref{parent}) will be short of the true SC phase coherence once there are some free $b$-spinons excited at $T\neq 0$. In this regime, $f$ should also vanish in the long-distance and long-time to lead to a finite broadening via a convolution with $\mathcal{D}_{a}$ in Eq. (\ref{Eq:3}). Nonetheless, this broadening effect is expected to be weak at low-temperature and should not change the basic conclusions to be drawn in the following. Thus, for the simplicity of presentation, we shall simply keep $F_0$ as a constant in Eq.(\ref{GPG}).

%is short but still larger than one lattice constant, indicating that most of the background $b$-spinons at low temperature stay in short-range singlet pairing states so that the $\pm$ vortices attached by $b$-spinons are almost canceled each other. Therefore, according to the Hamiltonian \eqnref{Eq:Hh}, holons $h_i$ now have coherence and condense, i.e., $\left\langle h_{i}\right\rangle \neq 0$, since gauge field $A_{ij}^{s}$ has little fluctuation, so that the behavior of ``twisted'' hole $\tilde{c}_i$ is mainly determined by that of backflow $a$-spinons.  Thus,  Then, in normal PG phase, the leading term of single-particle Green's function \eqnref{Eq:3} can be reduced to the form as: $G_{c, 0}(\boldsymbol{r}_{i}-\boldsymbol{r}_{j}, \tau ; \sigma)= F_0 D_{\tilde{c}}(\boldsymbol{r}_{i}-\boldsymbol{r}_{j}, \tau ; \sigma)$, and the general Dyson equation \eqnref{Eq:6} can also be expressed as a more compact form in momentum and frequency space:
For a free-electron band model, the single-particle Green's function in the normal state is given by
\begin{equation}
	\label{GC}
  G^c_{0}(\boldsymbol k;\omega)=\frac{1}{\omega-\epsilon_{k}^c}
\end{equation}
where $\epsilon_{k}^c=-2 t_{\mathrm{eff}}\left(\cos k_{x}+\cos k_{y}\right)-4 t_{\mathrm{eff}}^{\prime} \cos k_{x} \cos k_{y}+\mu$ is the dispersion for the cuprate with a large Fermi surface satisfying the Luttinger volume. In the following calculation, we take $t_{\text{eff}}=J$ and $t_{\mathrm{eff}}^{\prime}=-0.25t_{\mathrm{eff}}$, with the same set of the parameters given in Ref. \onlinecite{Weng.Ma.2014} which determine $\delta^*\simeq 0.26$ in the phase diagram of Fig. \ref{fig:phase_diagram}.

Therefore, we arrive at the single-particle Green's function in the LPP state at the RPA level by
\begin{equation}\label{GLPG}
  G^e(\boldsymbol k,\omega)=\frac{F_0}{\omega+\xi_{k+k_{0}}^a-\frac{\Delta_{ k+k_{0}}^{2}}{\omega-\xi_{k+k_{0}}^a}-\frac{\lambda^{2}F_0}{\omega-\epsilon_{k}^c}}.
\end{equation}

In \figref{fig:arcevo}(a)-(c), the quasiparticle spectral function $A(\boldsymbol k,\omega)=-\frac{1}{\pi} \Im G^e(\boldsymbol k,\omega)$ is shown at $\omega=0$,
%which is defined in terms of Eq. (\ref{GLPG}) by
%\begin{equation}
%  \label{Ak0}
 % A^e(k,\omega)=-\frac{1}{\pi} \Im G^e(k,\omega).
%\end{equation}
which gives rise to a Fermi arc structure at $\delta<\delta^*$.  The evolution trend as a function of doping indicates that the segment size of each Fermi arc enlarges monotonically as the doping level increases. The full Fermi surface will be eventually recovered at $\delta>\delta^*$ as shown in \figref{fig:arcevo}(d), which is to be discussed in \secref{sec5} below. 

\begin{figure*}[th]
	\includegraphics[width=\linewidth]{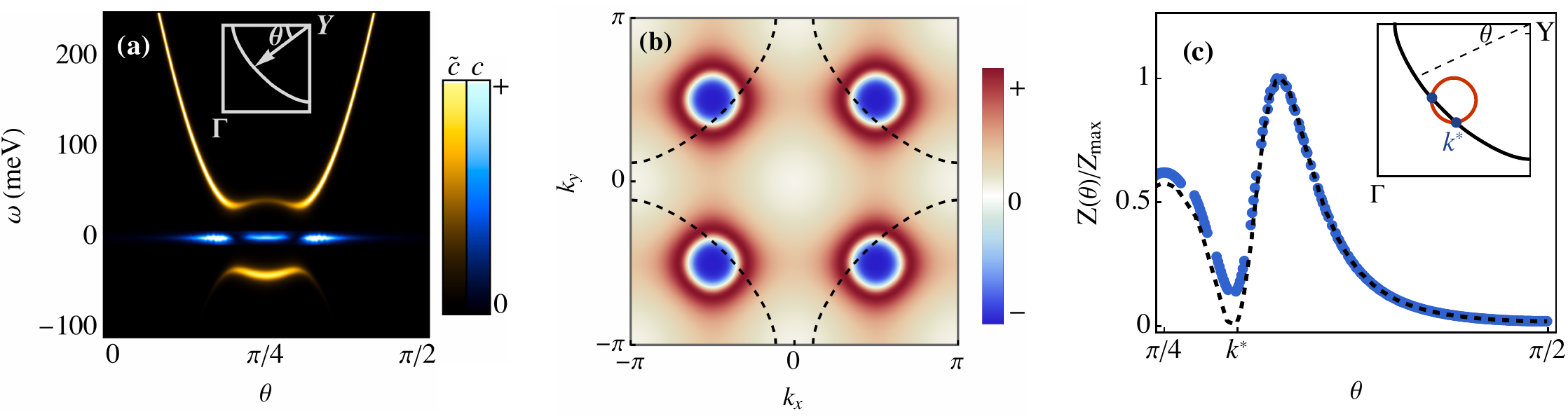}
	\caption{(a) The spectral weight (yellow) of the ``twisted'' hole depicted by $D_{\tilde c}(\boldsymbol k,\omega)$ along the contour with the angle $\theta$ marked in the inset. The quasiparticle (blue) emerges within the gap; (b) The real part of $D_{\tilde c}(\omega=0)$ which changes sign at the Fermi pockets of the $a$-spinons. The black dashed curves denote the large Fermi surface of the bare hole, i.e., $\epsilon_{k}^c=0$; (c) The spectral weight of the quasiparticle vs. angle $\theta$ with a vanishing (finite) imaginary broadening (see text) for the ``twisted'' hole: $\eta = 0 \text{meV}$ (black dashed) and $\eta = 2 \text{meV}$ (blue dot) under the replacement of $\omega \rightarrow \omega + i \eta $ in \eqnref{GLPG}. The inset of (c): the bare quasiparticle Fermi surface (black line) and the $a$-spinon pockets (red circle) at $\Delta^a=0$ intersects at two points labeled by the dark blue point at $k^*$. The doping concentration is at $\delta=0.1$.
	}
	\label{a-real}
\end{figure*}

\subsubsection{Nature of the ``Fermi arc''}\label{sec3A1}

The quasiparticle dispersion is determined by the zeros of the denominator in \eqnref{GLPG} at $\omega=E_{k,l}^e$ with
\begin{equation}\label{El}
  E_{k,l}^e=\frac{1}{3}\left[\epsilon_{k}^c+\operatorname{Re} e^{i 2l \pi / 3} (C_k+\sqrt{D_k})^{1 / 3}\right]
\end{equation}
which is composed of three branches: $l=0,1,2$. Here $C_k=-72 \Delta_{k}^{2} \epsilon_{k}^c+4(\epsilon_{k}^c-3 \xi_{k+k_{0}}^a)[9 \lambda^{2}F_0+2 \epsilon_{ k}^c(\epsilon_{ k}^c+3 \xi_{ k+k_{0}}^a)]$, $D_k=C_k^{2}-64[3 \Delta_{ k+k_{0}}^{2}+{\epsilon_{ k}^{c}}^2+3(\lambda^{2}F_0+{\xi_{ k+k_{0}}^{a}}^2)]^{3}$. Two branches of $E_{k,l}^e$ with negative values are shown in \figref{singularity}(a) in the $k_x$-$k_y$ plane (at $\delta=0.06$), while the third branch, being positive, is not shown there.

A ``pocket"-like branch of $E_{k,l}^e$ shown in \figref{singularity}(a) is at a finite energy below the Fermi energy, which is to be discussed later. Near the Fermi energy there is only one branch left, which is quite similar to the bare dispersion $\epsilon_{ k}^c$ with a full Fermi surface at $\omega=0$ in \figref{singularity}(a). However, the corresponding spectral function $A(\boldsymbol k,\omega=0)$ [cf. the top in \figref{singularity}(a)] only exhibits the Fermi arcs instead. In contrast, a full Fermi surface is restored at $\delta>\delta^*$ in \figref{singularity}(b) as to be discussed in \secref{sec5}.

Let us examine the quasiparticle spectral weight defined by
\begin{eqnarray}\label{weight}
    Z_{k,l}&=&\lim _{\omega \rightarrow E_{k,l}^e} G^e(\boldsymbol k, \omega) \times\left(\omega-E_{k,l}^e\right) \notag\\
    &=&F_0 \frac{\left(E_{k,l}^e-\xi_{k}^a\right)\left(E_{k,l}^e-\epsilon^c_{ k}\right)}{\prod_{n \neq l}\left(E_{k,l}^e-E_{k,n}^e\right)}.
\end{eqnarray}
As shown in \figref{a-real}(c), the weight $Z$ along the Fermi surface contour determined by $E_{k,l}^e=0$ in the LPP regime is dramatically modified: it maintains a finite weight near the nodal region ($\theta=\pi/4$) but is substantially suppressed in the antinodal region. In other words, the quasiparticle excitation still satisfies the Luttinger volume, but the spectral weight gets substantially renormalized by the strong correlation effect to result in a Fermi arc structure as manifested in \figref{fig:arcevo}(a)-(c).

\subsubsection{Fermi arc emerging within the gapped Fermi pocket of the $a$-spinon excitation}\label{sec3A2}

The suppression of the quasiparticle spectral weight in the antinodal region of the large Fermi surface contour can be attributed to the general fractionalization of a bare hole as the leading term in Eqs.(\ref{Eq:3}) and (\ref{GPG}). The spectral weight of $G^e_0$ and its real part are shown in \figref{a-real}(a) and (b), respectively. Here four Fermi pockets of the $a$-spinon in \figref{a-real}(b) are centered at $(\pm \pi / 2, \pm \pi / 2)$ in the Brillouin zone where the dashed large Fermi surface is marked as the position of the pole of $G^c_0$ in Eq. (\ref{GC}). The $a$-spinons are in the $s$-wave-pairing such that its spectrum is gapped as shown by the bright yellow curve in \figref{a-real}(a) for one of four pockets.

Then the quasiparticle spectral function $A(\boldsymbol k,\omega)$ at the Fermi energy $\omega=0$ along the large Fermi surface determined by $E_{k,l}^e=0$ [cf. the inset of \figref{a-real}(a)], characterized by the spectral weight function $Z(\theta)$ in \figref{a-real}(c), is illustrated by the blue in \figref{a-real}(a).

Thus, in the fractionalized mean-field state, the $a$-spinons as fermions form four (gapped) Fermi pockets according to Eqs. (\ref{GPG}) and (\ref{GaPG}). The Landau-like quasiparticle as a collective mode emerges as a new pole of $G^{e}$ at the RPA level, when the denominator $1-\lambda^{2} G^e_0 G_0^c$ vanishes in Eq. (\ref{Eq:7}). Note that at $\omega=0$, the free fermion propagator $G_0^c$ in \eqnref{GC} changes sign and diverges at the large Fermi surface $\epsilon_{k}^c=0$, while the real part of $G^e_0$ is prominent mainly in the pocket regions marked by the deep colors in \figref{a-real}(b). It implies that the new Fermi surface should coincide with the bare one of $\epsilon_{k}^c=0$ outside the pocket regions, but with quickly diminished spectral weight $Z$. It results in a pseudogap feature in the spectral function $A(\boldsymbol k,\omega)$ at small $\omega$ or the Fermi arc at $\omega=0$. On the other hand, near the nodal region, the RPA correction of $1-\lambda^{2} G^e_0 G_0^c$ becomes prominent in Eq. (\ref{Eq:7}), which leads to a nontrivial Fermi arc structure to be further elaborated below.

%%%%%%%%%%%%%%%%%%%%%%%%%%%%%%%%%%%%%%%%%%%%%%%%%%%%%%%%%%%%%%%%%%%%%%%%%%%%%%%%%%%%%%%%%%%%%%
\begin{figure*}[th]
	\includegraphics[width=\linewidth]{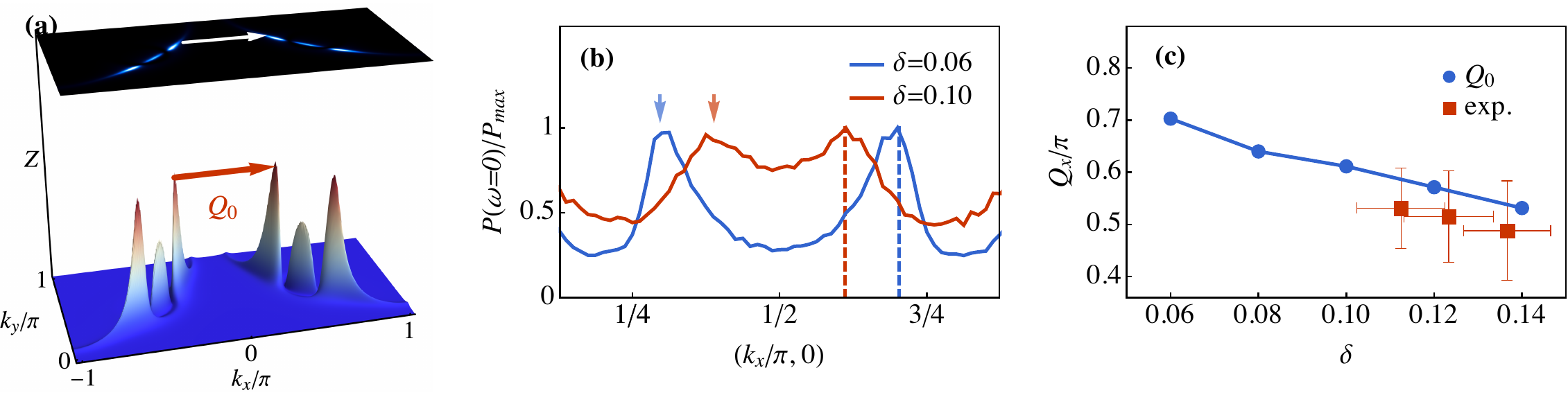}
	\caption{(a) The wavevector $Q_0$ connecting two ``hotspots'' with the largest weight $Z$; (b) The static QPI pattern obtained by \eqnref{QPIpattern} along $\left(k_{x}, 0\right)$ at $\delta=0.06$ (blue solid) and $\delta=0.10$ (red solid). The momentum $Q_0$ marked in (a) is indicated at the dashed vertical line at the right-hand-side peak of the QPI. Note that another peak in the QPI pattern appears approximately symmetric about $(\pi/2,0)$ as indicated by an arrow, whose origin is different and may not be intrinsic as argued in \appref{sec_App_QPI}; (c) Comparison of the momentum $Q_0$ with the CDW wavevector determined in the experimental measurement \cite{Damascelli.Comin.2014}.
	}
	\label{fig_CDW}
\end{figure*}

\subsubsection{Novel feature: breakpoints and hotspots}\label{sec3A3}

Each segment of the Fermi arc is actually broken into three pieces in \figref{fig:arcevo}(a)-(c), which correspond to the maximums of the quasiparticle weight $Z(\theta)$ along the large Fermi surface as \figref{a-real}(c) indicates. Here the maximum weight around the nodal region (with $\theta\simeq \pi/4$) is first quickly reduced around $k^*$ and then is back to the second sharp peak, called ``hotspot", at the edge of the Fermi arc, beyond which $Z(\theta)$ is finally monotonically diminished towards the antinodal region.

The ``breakpoint'' $k^*$ in \figref{a-real}(c) coincides with the intersection point of the bare quasiparticle Fermi surface with the $a$-spinon Fermi pocket (cf. the inset) where $\operatorname{Re} \mathcal D^a $ changes sign [indicated by the white circle in \figref{a-real}(b)]. Before a substantial reduction of $Z(\theta)$ towards to the antinodal region, the breakpoint $k^*$ is further accompanied by a quickly enhanced  ``hotspot'' in-between as illustrated in the main panel of \figref{a-real}(c).

It is noted that $Z(\theta)$ vanishes at $k^*$ may be just an artifact that the LPP is considered as the ground state in the present approach, instead of a finite-temperature phase above the superconducting phase at $T_c$. In Eq. (\ref{GLPG}), a finite $F_0$ as the long-range behavior of $f(\boldsymbol{r}_{i}-\boldsymbol{r}_{j}; \tau)$ is related to the phase coherence in the ground state. At finite temperature above $T_c$, the constant $F_0$ should be replaced by a convolution with $f(\boldsymbol{r}_{i}-\boldsymbol{r}_{j}; \tau)$ in Eq. (\ref{GLPG}) as thermally excited $b$-spinons will destroy the phase coherence. This effect may be minimally incorporated by setting a finite imaginary term in the denominator of the Green's function \eqnref{GaPG}, i.e., $\omega\rightarrow \omega+i\eta$, with the consequence  illustrated in \figref{a-real}(c) where the zero at $k^*$ is effectively lifted, although the dip is still clearly present.

Finally, it is noted that the above discussion on the Fermi arc is based on the ground state without SC coherence. In the SC state, however, the scalar RPA propagator in Eq. (\ref{Eq:7}) should be rewritten by using the Nambu matrix representation such that the Fermi arc will be replaced by a $d$-wave-like two-gap feature with the Landau quasiparticle becoming a Bogoliubov quasiparticle \cite{Weng.Zhang.2020}  (cf. \appref{sec_App_SC} for more discussions).

%%%%%%%%%%%%%%%%%%%%%%%%%%%%%%%%%%%%%%%%%%%%%%%%%%%%%%%%%%%%%%%%%

\subsubsection{CDW pattern}\label{secCDW}

The sharp peaks (``hotspots'') of the quasiparticle weight at the Fermi arcs will have important physical consequences. In the following, we investigate the static charge modulation due the quasiparticle interference (QPI) pattern at $\omega=0$ in the LPP.

For the electronic system with the single-particle Green's function \eqnref{GLPG}, a renormalized Green's function in the presence of impurity can be calculated via $t$-matrix method \cite{Lee.Wang.2003, Hirschfeld.Zhu.2004}. Here, the $t$-matrix describes a sequence of multiple scattering off the impurities:
\begin{eqnarray}
\hat{T}&=&\hat{V}+\hat{V} \hat G(\omega) \hat{V}+\hat{V} \hat G(\omega) \hat{V} \hat G(\omega) \hat{V}+\ldots \notag\\
&=&\hat{V}\left(1-\hat G(\omega) \hat{V}\right)^{-1}
\end{eqnarray}
where $\hat{V}_{i j}=v_{i} \delta_{i, j}$ and $\hat G(\omega)_{i j} = G^e(\boldsymbol r_i- \boldsymbol r_j, \omega)$ is the Green's function without impurity given by \eqnref{GLPG}. Here $v_{i}\sim \pm 0.05t$ denotes the local scattering potential, which changes signs randomly on randomly distributed sites with concentration $c\sim 5\%$. Then, the single-particle Green's function under the local impurity scattering can be expressed in the real space as:
\begin{eqnarray}
	G^{e\prime}(\boldsymbol r_i, \boldsymbol r_j , \omega) &= &G^e (\boldsymbol r_i- \boldsymbol r_j ; \omega)\\
	&\;&+\sum_{I J} G^e(\boldsymbol r_i- \boldsymbol r_I , \omega) T_{I J} G^e(\boldsymbol r_J- \boldsymbol r_j , \omega)\notag
\end{eqnarray}
The local density of states (LDOS) at a general site $i$ can be decomposed as $\rho(\boldsymbol r_i,\omega)=\rho_{0}(\omega)+\delta\rho(\boldsymbol r_i,\omega)$, where  $\rho_{0}$ is the homogeneous density of state, and $\delta\rho$ is the local shift due to disorder. Here, $\delta\rho$ is determined by the analytic continuation $G^{e\prime}\left(\boldsymbol r_i, \boldsymbol r_j ; i \omega_{n}\right) \rightarrow G^{e\prime}\left(\boldsymbol r_i, \boldsymbol r_j ; \omega+i 0^{+}\right)$:
\begin{equation}\label{LDOS}
	\delta\rho(\boldsymbol r_i,\omega)=-\frac{1}{\pi} \operatorname{Im} G^{e\prime}\left(\boldsymbol r_i, \boldsymbol r_i ; \omega+i 0^{+}\right)-\rho_{0}(\omega)
\end{equation}
Finally, for a certain energy $\omega$, the QPI pattern in momentum space is calculated as the power spectrum of the LDOS:
\begin{equation}\label{QPIpattern}
	P(\boldsymbol{q}, \omega)=\left|\frac{1}{N} \sum_{i} \delta\rho(\boldsymbol r_i,\omega) e^{-i \boldsymbol{q} \cdot \boldsymbol{r}_{i}}\right|^{2}
\end{equation}
where $\boldsymbol{q}$ is the QPI momentum.

\begin{figure*}[th]
	\includegraphics[scale=0.95]{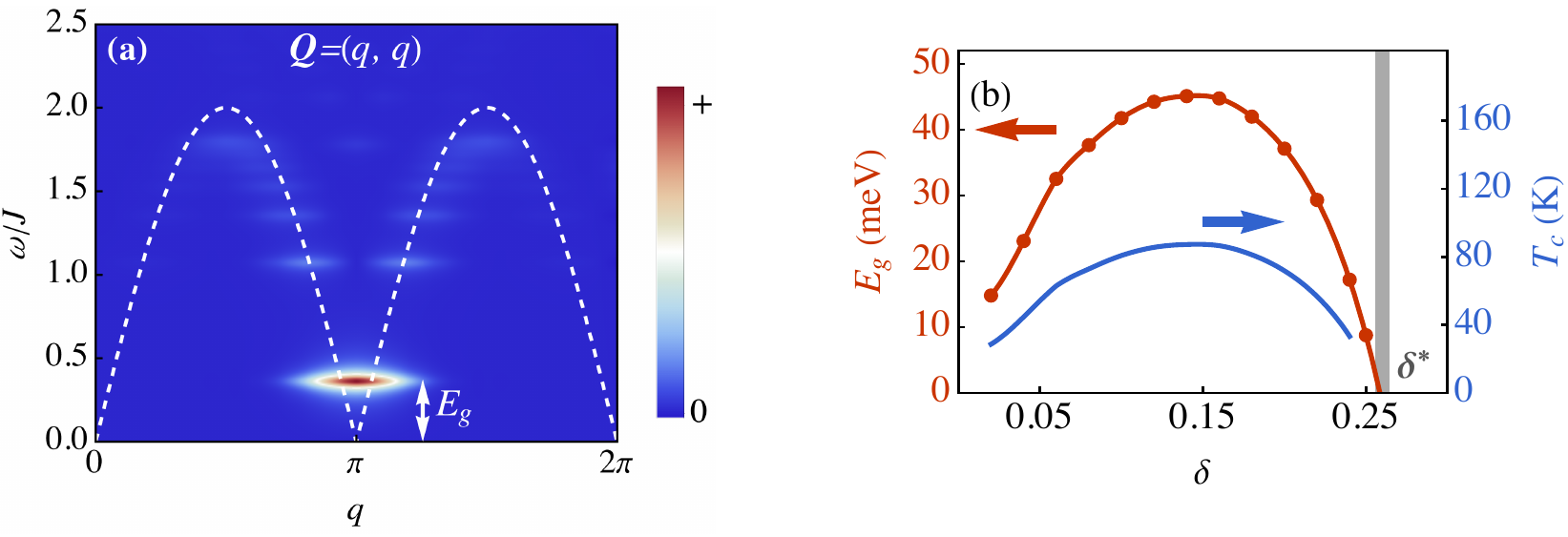}
	\caption{(a) Imaginary part of dynamic spin susceptibility obtained by \eqnref{chib} along the diagonal line at $\delta=0.1$. The characteristic energy of a sharp low-lying resonance mode is labeled by $E_g$, and the white dashed line denotes the spin-wave dispersion at half-filling case. (b) The doping evolution of $E_g$ (red solid dot) obtained by mean-field self-consistent calculations in \appref{spin} and corresponding SC phase transition temperate $T_c$ (blue) according to \eqnref{Tc}. As doping density tends to critical point $\delta^*$, $E_{g}$ also reduces to zero.
	}
	\label{fig_Eg}
\end{figure*}

In the LPP, according to \eqnref{GLPG}, the unconventional single-particles Green's function $G^e$ in the absence of impurity shows that there are two isolated “hotspots” with conspicuous enhancement of the density of states for each segment of Fermi arcs. The nontrivial QPI pattern can be induced by the scattering process induced by impurity between these ``hotspots". Here, the calculated QPI patterns are most prominent along $(k_x,0)$ as illustrated in \figref{fig_CDW}(b) at various doping concentrations, in which the intensity peaks are marked by the dashed vertical lines.  Each of this peak coincides with the momentum $Q_0$ connecting  two \emph{outer} ``hotspots'' of the neighboring Fermi arcs as indicated in \figref{fig_CDW}(a). In addition, the QPI pattern in \figref{fig_CDW}(b) appears to be symmetric about $(\pi/2,0)$, with another peak labeled by the arrow. But such a symmetric pattern may happen to arise due to a pseudo-nested structure in the single-particle Green's function \eqnref{GLPG} under the current parameter setting and the additional peak (labeled by the arrow) may not be robust, which can be reduced as a slightly changed structure is considered. More discussion about this symmetric peak is given in \appref{sec_App_QPI}.

The momentum $Q_0$ varies as a function of doping as summarized in \figref{fig_CDW}(c). It shows that $Q_0$ decreases monotonically with doping, which is qualitatively in agreement with the experimental measurements of the charge-density modulations in the cuprates \cite{Damascelli.Comin.2014, Davis.Fujita.2014, Hudson.Wise.2008, Ghiringhelli.Peng.2016}. A common feature of the QPI patterns at various doping concentrations is that the characteristic momentum coincides with the momentum transfer between two specific “hotspots” of the Fermi arcs shown in \figref{fig_CDW}(a).

Note that here we do not consider the charge modulation pattern as a conventional CDW order that spontaneously breaks translational symmetry in the LPP. Instead, this modulation is simply a byproduct of the presence of a Fermi arc ``hot spot", namely a QPI pattern induced by scattering via impurities. In addition, the relationship between the charge modulation and the SC state is clear: the effect of charge modulation is irrelevant for the establishment of the SC phase coherence. But the single-particle spectrum will open a ``V-shaped" gap in the SC phase \cite{Weng.Zhang.2020}, leaving the ``hot spot'' feature diminishing with the Fermi arc, which substantially weakens the intensity of the QPI pattern in consistency with the experimental observations \cite{Hayden.Chang.2012, Braicovich.Ghiringhelli.20120wb}.

	Finally, it is worth noting that the observed charge modulation results in STM\cite{Damascelli.Comin.2014, Hudson.Wise.2008, Davis.Fujita.2014} can be explicitly identified in the QPI pattern presented in this section. On the other hand, the charge instability observed in other spectroscopic experimental results, such as several X-ray measurements\cite{Damascelli.Comin.2014, Ghiringhelli.Peng.2016, Hayden.Blackburn.2013, Hayden.Chang.2012, Braicovich.Ghiringhelli.20120wb}, can be directly associated with the dynamical charge susceptibility. It is important to recognize that both of them are manifestations of the momentum distribution of density at the Fermi surface within this theoretical framework, indicating that they are inherently consistent. Further discussions about the dynamical charge susceptibility are presented in the Appendix, which illustrates how the dynamical charge susceptibility function gets enhanced at $Q_0$ and may lead to a true charge-density-wave instability.

%%%%%%%%%%%%%%%%%%%%%%%%%%%%%%%%%%%%%%%%%%%%%%%%%%%%%%%%%%%%%%%%
\subsection{Restoring full Fermi surface at $\delta>\delta^*$}\label{sec3B}

The Fermi arc for the gapless Landau-like quasiparticle excitation has been established in the LPP.  In the fractionalized mean-field state of Eq. (\ref{parent}), the bosonic RVB order parameter $\Delta^s$ will vanish beyond a critical doping concentration, to be denoted by $\delta^*$, which defines the boundary of the LPP in the overdoping. In the following, we shall examine the single-particle Green's function at the RPA level in Eq. (\ref{Eq:7}) and inspect how the corresponding Fermi surface will evolve into $ \delta>\delta^*$.

%Beyond $\delta^*$, the fractionalized sub-states in Eq. (\ref{parent})

\subsubsection{Vanishing bosonic-RVB order at $\delta \rightarrow \delta^*$}\label{sec3B1}

Since $\Delta^s$ is the main controlling order parameter of the fractionalized state in Eq. (\ref{parent}), in the following, we first examine the basic spin dynamics of the $b$-spinons at the mean-field level $\Delta^s\neq 0$ before taking the limit $\Delta^s= 0$   in $|\Phi_b\rangle$.

A typical dynamic spin susceptibility $\chi ^{\prime \prime}(\boldsymbol{q},\omega)$ calculated based on the ground state $|\Phi_b\rangle$ at $\delta=0.1$ by a standard mean-field procedure \cite{Ting.Weng.1999, Weng.Chen.2005}(cf. \appref{spin} for more details) is shown in \figref{fig_Eg}(a), in which the dispersion of the spin-wave obtained in the same theory at half-filling is plotted as the white dashed curve for comparison. Here the spin wave reduces to a low-lying mode at a ``resonance" energy $E_g$ around the AFM wavevector $\boldsymbol{Q}=(\pi,\pi)$ in the LPP (at $\delta=0.1$) in consistency with the neutron-scattering experiments \cite{Dai1999, Keimer.Fong.1999, Keimer.He.20006wc, Keimer.He.2002,Keimer.Capogna.2007, Bourges.Fauqu.2007}, which is continuously softened with reducing doping and eventually becomes gapless to recover the spin wave in the AFM phase. A detailed doping evolution of $E_g$ is shown in \figref{fig_Eg}(b) by the red curve, which is calculated under the same set of parameters used to determine the mean-field phase diagram in Fig. \ref{fig:phase_diagram} [cf. \appref{sec_App_selfconsis} for more calculation details].

As mentioned previously, the $b$-spinon excitations are crucial in disordering the superconducting phase coherence at finite temperature. As a matter of fact, a superconducting instability will occur in the LPP when the temperature is lower than \cite{Weng.Mei.20107w}
\begin{equation}\label{Tc}
T_c=\frac {E_g}{\kappa k_B}
\end{equation}
with $\kappa\simeq 6$, which is closed to the experimental consequences \cite{Dogan.Dai.1996, Aksay.Fong.1997, Keimer.He.20006wc, Colson.Gallais.2002}. Namely a finite $E_g$ will ensure a superconducting phase coherence by suppressing the free $b$-spinon excitations below $T_c$, which is also plotted in \figref{fig_Eg}(b). Both $E_g$ and $T_c$ will get reduced continuously in the overdoping regime as $\delta\rightarrow \delta^*$ with the suppression of the bosonic RVB order in the overdoping.

\begin{figure}[t]
	\centering
	\includegraphics[width=\linewidth]{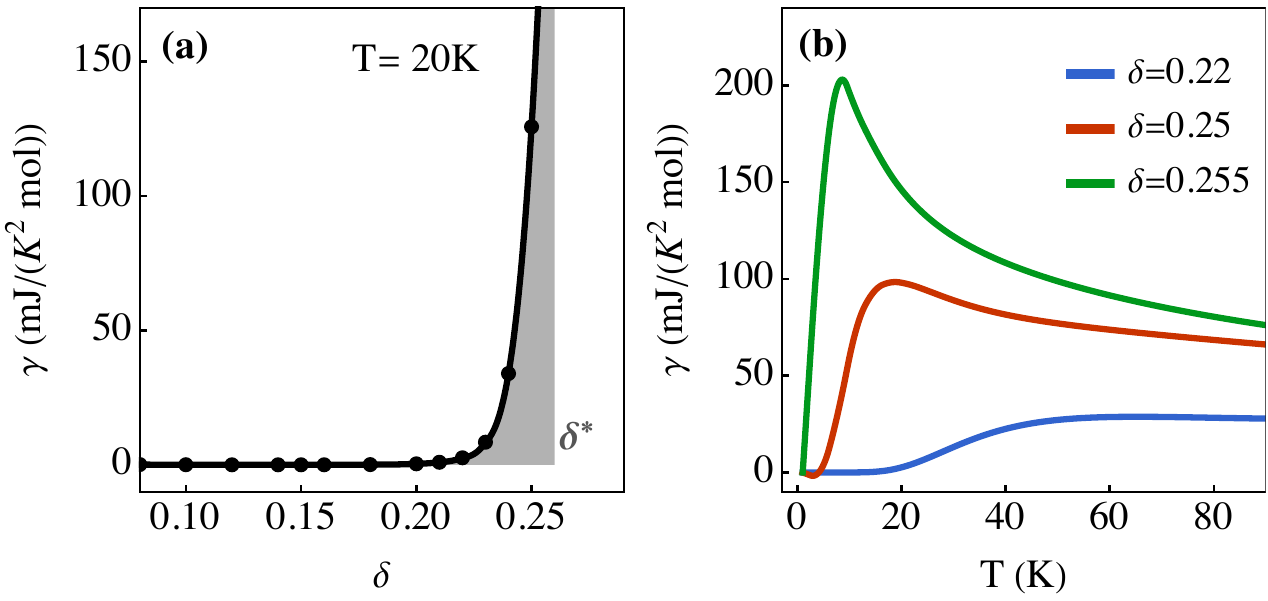}
	\caption{
		 (a) The doping evolution of specific heat $\gamma\equiv C_v/T$ contributed by the $b$-spinons as given in \eqnref{eqnCv} at $T=20K$, which exhibits a large peak as $\delta\rightarrow\delta^*$. (b) The temperature dependence of $\gamma$ at different dopings.
	}
	\label{fig_Cv}
\end{figure}

A characteristic manifestation of the change in the LPP is the specific heat $C_{v}^{b}$ contributed by the $b$-spinons. As shown in  \figref{fig_Cv}(a), a significant enhancement of $ \gamma \equiv C_{v}^{b} / T$ given in \eqnref{eqnCv} is found as the doping approaches $\delta^*$ at a given temperature, which is in agreement with the experimental results \cite{Taillefer.Proust.2019, Klein.Girod.2020}. A systematic evolution of $ \gamma$ vs. $T$ at different doping close to $\delta^*\simeq0.26$ is further illustrated in \figref{fig_Cv}(b).  Both of them indicate that a spike of the specific heat well reflects the fact that the spectrum of the $b$-spinons get strongly suppressed as the LPP ends at $\delta^*$.

Note that the origin of the huge peak of specific heat $\gamma$ at $\delta^*$ has been attributed \cite{Taillefer.Proust.2019,Klein.Girod.2020} in the literature to the divergence in the effective mass $m^*$ of the quasiparticles, according to the relation $\gamma \varpropto m^{*}$ in a Fermi liquid theory. However, in the present two-component RVB theory, the electronic structure of the quasiparticles changes smoothly around $\delta^*$ with $G_0^{c}$ remains the same in the Dyson equation \eqnref{Eq:6} except for the restoration of the spectral weight near the antinodal region. Here a sharp divergent specific heat at $T=0$ can only come from the background spin degrees of freedom as shown above.

Finally, it is noted that so far in this subsection only the contribution of the background $b$-spinons to the spin spectrum is considered, while the contribution of the backflow $a$-spinons is ignored. This is because the $a$-spinons behave more like the itinerant fermions with a larger energy gap and its contribution to the dynamic spin susceptibility is a gapped continuum at higher energies. At the mean-field level, the low-energy spin correlation is dominated by the RVB-pairing $b$-spinons originating from local moments, as depicted in \figref{fig_Eg}(a). The influence of $a$-spinons on this correlation has been examined at an RPA level in the other work\cite{Zhang.Weng.hg_2023}, which can shed further light on the high-energy spin excitations probed by resonant inelastic x-ray scattering (RIXS)\cite{Keimer.Tacon.2011, Dean.Meyers.2017, Ghiringhelli.Peng.2018ome, Zhou.Robarts.2021, Hayden.Robarts.2019} and provide an explanation for the detailed hour-glass feature \cite{Dogan.Hayden.2004,Yamada.Tranquada.2004, Fujita.Sato.2020} near $(\pi, \pi)$.

The influence of $a$-spinons on this correlation will be examined at an RPA level in future work, which can shed further light on the high-energy spin excitations probed by resonant inelastic x-ray scattering (RIXS)\cite{Keimer.Tacon.2011, Dean.Meyers.2017, Ghiringhelli.Peng.2018ome, Zhou.Robarts.2021, Hayden.Robarts.2019} and provide an explanation for the detailed hour-glass feature \cite{Dogan.Hayden.2004,Yamada.Tranquada.2004, Fujita.Sato.2020} near $(\pi, \pi)$.

%%%%%%%%%%%%%%%%%%%%%%%%%%%%%%%%%%%%%%%%%%%%%%%%%%%%%%%%%%%%%%%%%%%%%%%%%%%%%%%%%%%%%%%%%%%%%%

%%%%%%%%%%%%%%%%%%%%%%%%%%%%%%%%%%%%%%%%%%%%%%%%%%%%%%%%%%%%%%%%%%%%%%%%%%%%%%%%%%%%%%%%%%%%%%
\subsubsection{Full Fermi surface at $\delta > \delta^*$}\label{sec5}

With the RVB order parameter $\Delta^s=0$ at $\delta > \delta^*$, the background $b$-spinons are completely localized in $|\Phi_b\rangle $ with $\xi\rightarrow 0$. A classical Curie-Weiss behavior of the local moments is expected at finite temperatures in such a phase with $\Delta^s=0$ \cite{Weng.Ma.2014}. A direct consequence is that the phase in Eq. (\ref{f}) is totally spatially disordered such that
\begin{equation}\label{f1}
f(\boldsymbol{r}_{i}-\boldsymbol{r}_{j}; \tau) \simeq (2\pi)^2\delta(\boldsymbol{r}_{i}-\boldsymbol{r}_{j}).
\end{equation}
This is a duality behavior of the phase-string factor in the so-called strange-metal phase characterized by $\Delta^s=0$.

As the result, the fractionalized mean-field propagator of \eqref{Eq:3} is reduced to be local in space such that featureless in $\boldsymbol k$:
\begin{equation}
G^e_0(\boldsymbol k, \omega)\rightarrow \Sigma_{\tilde{c}}( \omega)\propto \frac 1 {L^2} \sum_{\boldsymbol q}D_{\tilde{c}}(\boldsymbol q, \omega)
\end{equation}
which means the twisted quasiparticle $\tilde{c}$ defined in Eq. (\ref{ctilde}) can no longer coherently propagate in the totally disordered spin background. Consequently, the Dyson equation \eqnref{Eq:6} at the RPA level can be organized as follows:
\begin{eqnarray}\label{GFL}
    G^e(\boldsymbol k, \omega)
    %&=& \Sigma_{\tilde{c}}+\lambda^{2} \Sigma_{\tilde{c}} G^c_{0}(\boldsymbol k, \omega) \Sigma_{\tilde{c}}+\ldots \notag \\
		&=&\Sigma_{\tilde{c}}(\omega)+ \lambda^{2} \Sigma_{\tilde{c}}^{2}(\omega)\left[G^c_{0}(\boldsymbol k, \omega)\right. \notag \\
		& &  \ \left. +\lambda^{2} G^c_{0}(\boldsymbol k, \omega)  \Sigma_{\tilde{c}}(\omega) G^c_{0}(\boldsymbol k, \omega) + \ldots \right]  \notag \\
    &=& \Sigma_{\tilde{c}}(\omega)+\frac{\lambda^{2} \Sigma_{\tilde{c}}^{2}(\omega)}{{G^c_0}^{-1}(\boldsymbol k, \omega)-\lambda^{2} \Sigma_{\tilde c}(\omega)} \notag \\
    &=& \Sigma_{\tilde{c}}(\omega)+\frac{\lambda^{2} \Sigma_{\tilde{c}}^{2}(\omega)}{\omega-\epsilon_k^c-\lambda^{2} \Sigma_{\tilde c}(\omega)}.
\end{eqnarray}

Therefore, by a simple shift of the chemical potential in $\epsilon_k^c$ in the denominator of the second term in Eq. (\ref{GFL}), the single-particle Green's function recovers the form of the non-interacting Fermi gas with an overall reduction of the spectral weight by a factor of $\lambda^{2} \Sigma_{\tilde{c}}^{2}$, besides a trivial first term. The corresponding large Fermi surface satisfying the Luttinger volume is shown in Fig. \ref{fig:arcevo}(d) in terms of the corresponding spectral function at $\omega=0$. Namely, as far as the quasiparticle excitation is concerned, a $T=0$ phase at $\Delta^s=0$ looks like a Fermi liquid with a large Fermi surface.

However, it is important to note that even though the single-particle Green's function at $T=0$ describes the coherent gapless quasiparticle mode near the large Fermi surface in Eq. (\ref{GFL}), the ``strange metal'' phase at $\delta>\delta^*$ is not simply a Fermi liquid state. We have pointed out that the background $b$-spinons are totally decoherence at $\delta>\delta^*$, which in turn lead to the phase incoherence in Eq. (\ref{f1}) such that the single-particle excitation is effectively decoupled from the dynamics of the fractionalized particle $\tilde{c}$, except for an overall renormalization factor $\Sigma_{\tilde{c}}$ in Eq. (\ref{GFL}). In other words, the fractionalized particles can still play an important role especially in the finite-temperature regime in such a phase \cite{Weng.Weng.2011,Weng.Ma.2014}, where the overall weight of the quasiparticle is further reduced by an amount of $\lambda^{2} \Sigma_{\tilde{c}}^{2}$ in Eq. (\ref{GFL}) at $T=0$, whose physical meaning will be further discussed in the next section.

\begin{figure}[t]
	\centering
	\includegraphics[scale=0.35]{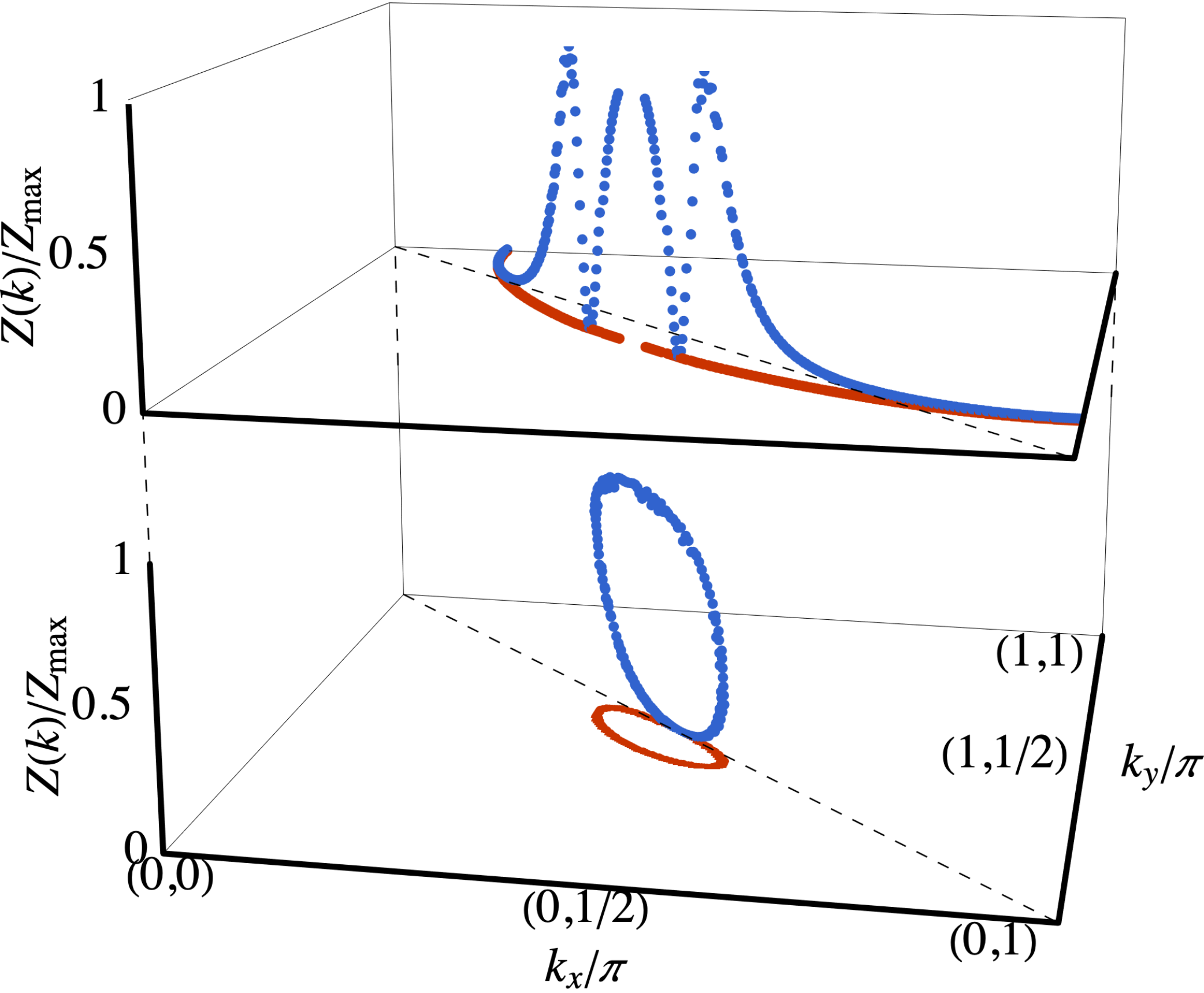}
	\caption{
		The renormalized spectral weight (blue dot) and the pole or Fermi surface (red curve) of the single-particle Green's function in this paper (upper panel) and in the YRZ theory \cite{Zhang.Yang.2006} (bottom panel). The dashed line denotes the AFM Brillouin zone boundary.
	}
	\label{YRZ}
\end{figure}

%%%%%%%%%%%%%%%%%%%%%%%%%%%%%%%%%%%%%%%%%%%%%%%%%%%%%%%%%%%%%%%%%%%%%%%%%%%%%%%%%%%%%%%%%%%%%%

\section{Discussion}\label{sec6}

As a basic characteristic of the pseudogap phase in the cuprate, the origin of the Fermi arc observed in the ARPES experiment has been an intensive focus of theoretical studies. In the following, we shall further discuss the underlying physics for the Fermi arc and pseudogap structure in the present approach based on the analytic structure of the single-particle Green's function in Eq. (\ref{GLPG}).  A comparison with other approaches will be also made to show the important distinctions, which concern the nature of characterizations of strong correlation effect at $\delta<\delta^*$ and beyond.

\subsection{Large Fermi surface vs. small Fermi pocket}

First of all, as shown in the top panel of Fig. \ref{YRZ}, the pole of Eq. (\ref{GLPG}), i.e., the dispersion given in Eq. (\ref{El}) in the first quadrant of the Brillouin zone, does give rise to a full large Fermi surface at $\omega=0$ (red curve) which still satisfies the \emph{large} Luttinger volume consistent with Oshikawa's topological argument \cite{Oshikawa.Oshikawa.2000} for a quasiparticle excitation. On the other hand, it is the quasiparticle spectral weight $Z(k)$ on such a Fermi surface that exhibits a strong unconventional suppression towards the anti-nodal points, resulting in a Fermi-arc-like structure shown in \figref{fig:arcevo}. In particular, the edges of the ``Fermi arc'' are bordered by two sharp peaks (the hotspots), which are separated, by two sharp dips, from the central portion of the nodal region in the top panel of Fig. \ref{YRZ}. Here the dips are at the intersections between the large Fermi surface with a Fermi pocket of the $a$-spinon, which is gapped as indicated in \figref{a-real} (a) and (b), wherein the quasiparticle is well protected with a significant weight. Outside such a ``Fermi arc'', the fractionalization of the Landau's quasiparticle into an $a$-spinon (the $h$-holon is condensed) [cf. Eq. (\ref{Eq:1})] becomes predominant, which significantly weakens the quasiparticle weight. In a phenomenological theory of the pseudogap proposed by Senthil and Lee \cite{Lee.Senthil.2009}, a similar picture that the quasiparticle still retains a portion of the large Fermi surface near the nodal region has been discussed, with a strongly anisotropic spectral weight being attributed to the scattering by the fluctuations of the $d$-wave SC pairing field\cite{Lee.Senthil.2009, Norman.Micklitz.2009}. In the present microscopic approach, however, the strong reduction of the spectral weight near the antinodal region is quantitatively realized by a novel fractionalization process, where the pairing of the $a$-spinons does serve as the amplitude of the SC order parameter in Eq. (\ref{SC}).

An alternative Fermi arc structure is shown in the bottom panel of Fig. \ref{YRZ} based on a phenomenological description of the single-particle Green's function, which has been previously proposed by Yang, Rice, and Zhang, known as the YRZ theory \cite{Zhang.Yang.2006}, based on the Hubbard model. It is given as follows:
\begin{equation}\label{gYRZ}
  G^{\mathrm{YRZ}}(\boldsymbol k, \omega)=\frac{g_{t}}{\omega-\epsilon^c_k-\Delta_{R}^{2} /\left(\omega+\xi_{0, k}\right)}
\end{equation}
where $\Delta_{R}$ denotes a $d$-wave RVB order parameter and $\xi_{0, k}$ the dispersion of the fermionic spinons in the spirit of the original Anderson's (one-component) RVB theory. In the bottom panel of \figref{YRZ}, the corresponding pole at $\omega=0$ and the spectral weight along the Fermi surface are shown in the first quadrant of the Brillouin zone. It clearly indicates that the ``Fermi arc'' appearing here is actually originated from an underlying small Fermi pocket at $\omega=0$, which does not obey the large Luttinger volume until in overdoping where the RVB gap $\Delta_{R}$ vanishes \cite{Zhang.Yang.2006}. As illustrated in the bottom panel of \figref{YRZ}, the ``Fermi arc'' here is due to the suppression of the spectral weight $Z(k)$ in the half portion of the Fermi pocket which comes in contact with the AFM Brillouin zone boundary [the dashed line in Fig. \ref{YRZ}].

Based on the spin-fermion model, Qi and Sachdev \cite{Sachdev.Qi.2010} also treat a quasiparticle excitation as a bound state of a spinless fermion and a bosonic spinon with the binding force mediated by an emergent $U(1)$ gauge field. Here the quasiparticles form small Fermi pockets near the node region as the result of a reconstruction of the original large Fermi surface by the fluctuating AFM order, with opening up a pseudogap near the antinode region. Similar to the YRZ theory  \cite{Zhang.Yang.2006}, it has been argued \cite{Sachdev.Qi.2010} that the quasiparticle weight for the outer half of the pockets get suppressed to lead to a ``Fermi arc'' structure as observed in ARPES. 

\subsection{Origin of the pseudogap in the single-particle spectral function}

In this work, we have studied the Landau-like quasiparticle excitation emerging as a ``collective mode'' at the RPA level based on the fractionalized mean-field state described by Eq. (\ref{parent}). Indeed, according to the equation-of-motion study \cite{Weng.Zhang.2020}, a bare hole created by $\hat c_{k\sigma}$ on such a mean-field background can still propagate coherently with a renormalized band structure of $\epsilon^c_k$ before it decays into the fractionalized elementary particles. Here the quasiparticle excitation obeys the Luttinger volume with a large Fermi surface as discussed above. But its fractionalization as characterized by Eq. (\ref{Eq:1}) will strongly influence the behavior of the single-particle Green's function of Eq. (\ref{Eq:7}). The RPA equations in Eqs. (\ref{Dyeq}), (\ref{Eq:6}) and (\ref{Eq:7}) are determined by the interplay between the bare quasiparticle propagator $G^c_0$ [Eq. (\ref{GC})] and the fractionalized propagator $G^e_0$ [Eq. (\ref{Eq:3})] with a strength $\propto \lambda$.

\begin{figure}[t]
	\centering
	\includegraphics[scale=0.5]{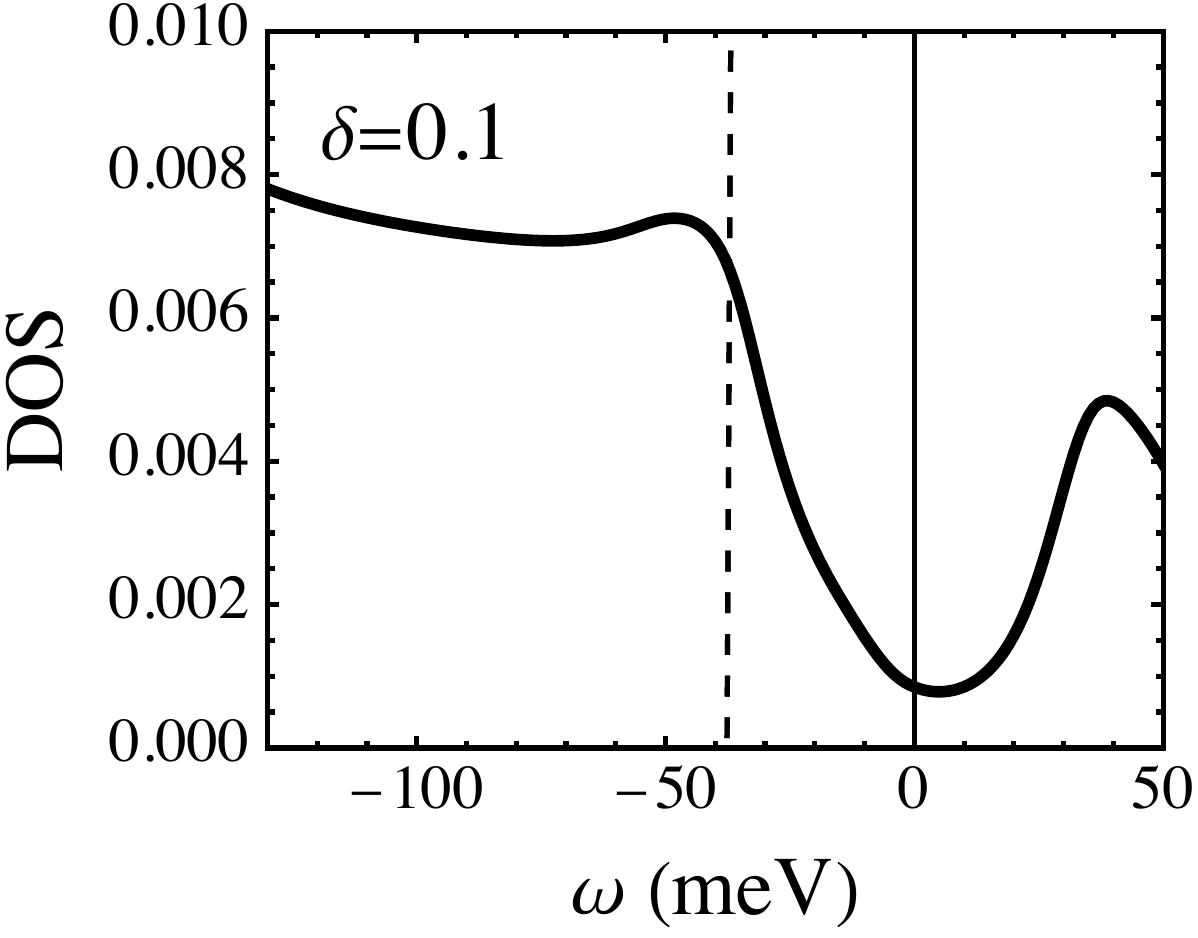}
	\caption{The DOS spectrum calculated by integrating $G^c$ over the entire BZ at $\delta=0.1$. Here the lifetime of ``twisted'' quasiparticle $\tilde c$ coming from the drift of free $b$-spinons takes the value of $\varsigma=10$meV when taking $\omega \rightarrow \omega + i \varsigma$ in \eqnref{Eq:4}. The dashed lines label the $a$-spinon gap.}
	 	\label{fig:STM}
\end{figure}

The pseudogap behavior in the single-particle spectral function is thus clearly associated with the gap opening due to the $a$-spinon pairing in the fractionalized propagator $G^e_0$ as illustrated in Fig. \ref{a-real}(a). 
	Additionally, a more explicit connection can be uncovered through the investigation of the local DOS spectrum in \figref{fig:STM}, obtained by integrating out the momentum in quasiparticle spectral function, i.e., $\sum_{k}A_c(k,\omega)$. Clearly the reduced but finite weight at low $\omega$ (as dictated by $\Delta_a$ labeled by the dashed line in \figref{fig:STM}) is consistent with the particle-hole asymmetry pseudogap feature observed in STM experiments\cite{Fischer.Renner.1998,Hoffman.He.2014}. It should be noted that the structure of the phase correlation function in \eqnref{f} at finite frequency remains to be discussed, as it may give rise to some high-energy spectroscopic details through the convolution in the Dyson equation \eqnref{Eq:6}. However, for the sake of simplicity in this presentation, we assume that the effect of drift for free $b$-spinon is to give the ``twisted'' hole $\tilde c$ a finite lifetime of $\varsigma=10$ meV through the replacement $\omega \rightarrow \omega + i \varsigma$ in \eqnref{Eq:4}.

Within the region covered by the gapped small Fermi pocket of the $a$-spinons [cf. Figs. \ref{a-real}(a) and (b)], the bare quasiparticle remains decoupled from the fractionalization to maintain a coherent Fermi arc, whereas a strong suppression of its spectral weight outside the $a$-pocket is due to the process of a strongly enhanced decay into the fractionalized objects. The high-energy spectral weight of $G^e_0$ above the $a$-spinon gap will then dominate such a region which extends over to the antinodal region outside the ``Fermi arc'', responsible for the pseudogap feature in the spectral function. %But such a ``gap'' is not truly associated with the quasiparticle excitation with $Z(k)\sim 0$ in this region.

Therefore, the pseudogap phenomenon in the LPP is a direct manifestation of the electron fractionalization of Eq. (\ref{Eq:1}), which is characterized by Eqs. (\ref{SC}) and (\ref{phdisor}) with $\Delta^a\neq 0$. The origin of the Fermi arcs in Figs. \ref{fig:arcevo}(a)-(c) at $\delta<\delta^*$ is the natural consequence of such a fractionalization without violating the Luttinger volume, which then naturally evolves into a manifest large Fermi surface at $\delta>\delta^*$ with $\Delta^s=0$. In the latter case, the uniformly reduced quasiparticle weight $\lambda^{2} \Sigma_{\tilde{c}}^{2}$ in Eq. (\ref{GFL}) also hints at the incoherent components hidden behind the conventional ``Fermi liquid'', which are the remaining traces of fractionalized particles in the overdoping.

Finally, note that the primary focus of this research is to investigate the lower pseudogap phase, which terminates at $\delta^*$, beyond which the system exhibits behavior consistent with that of a Fermi liquid. Such critical point at $\delta^*$ is characterized by a specific heat divergence at low temperatures, which is consistent with experimental observations\cite{Taillefer.Proust.2019, Klein.Girod.2020}. Notably, this study does not focus on superconductivity. Previous research, such as that presented in \cite{Weng.Ma.2014}, has demonstrated that the superconducting instability at $T_c$ concludes at $\delta^*$, as depicted in \figref{fig_Eg}(b). However, it is plausible that $T_c$ may extend to higher doping concentrations beyond $\delta^*$, as the Fermi liquid phase may not remain stable against further superconducting condensation. Thus, while the phase diagram presented in \figref{fig:phase_diagram} reflects the mean-field level results obtained in \cite{Weng.Ma.2014}, it cannot definitively exclude the possibility of the endpoints of the superconducting phase extending beyond $\delta^*$\cite{Taillefer.Proust.2019}.

\subsection{Small Fermi pockets emerging at $\Delta^a=0$}

\begin{figure}[th]
	\centering
	\includegraphics[width=\linewidth]{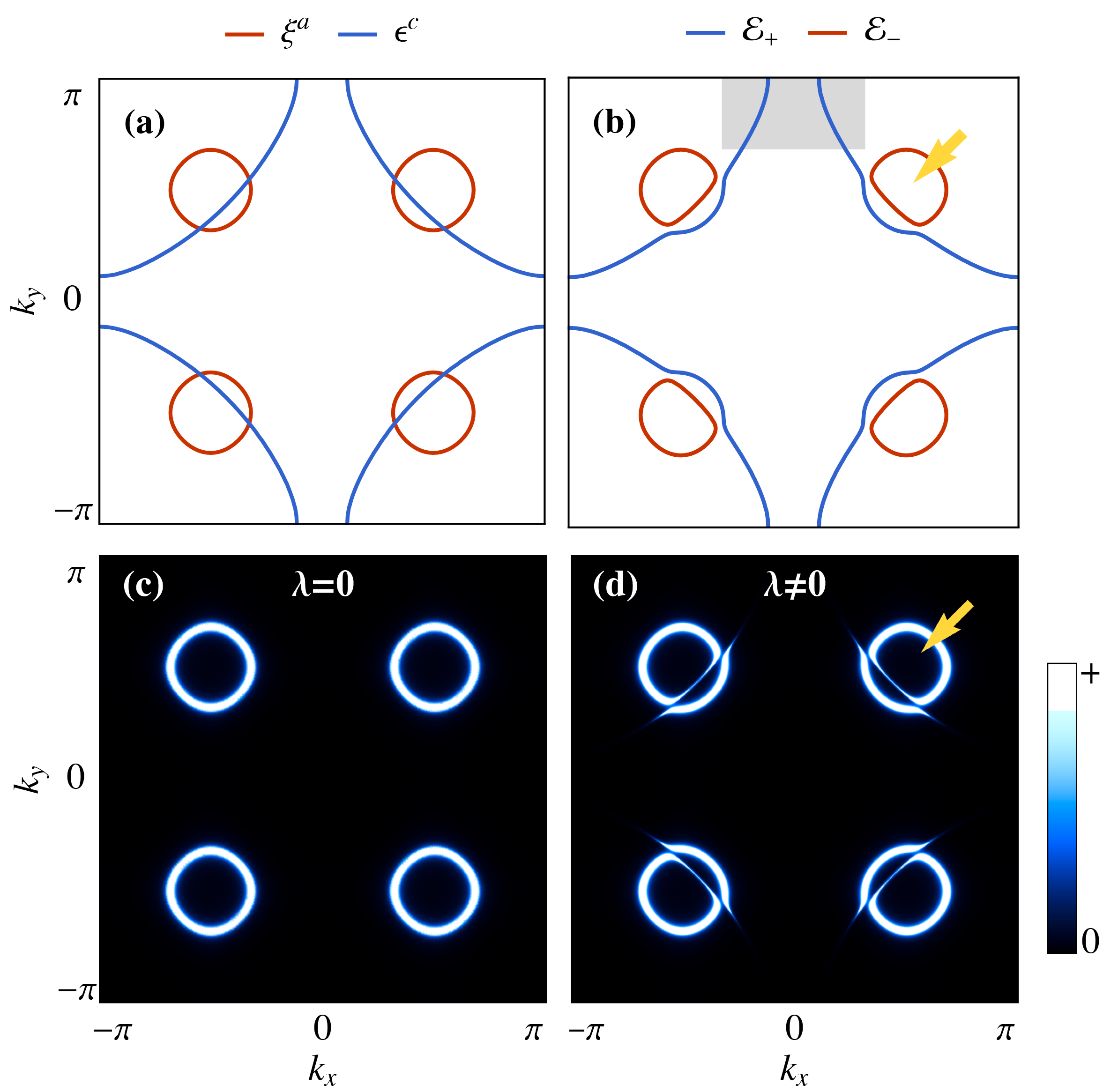}
	\caption{The Fermi surfaces for $a$-spinons and bare holes when $\Delta^a=0$ are illustrated with $\lambda=0$ in (a) , and with $\lambda\neq 0$ in (b). Near the anti-nodal regions, as indicated by the gray region in (b), there is a suppression of spectral weight. The quasiparticle spectral function $A(k, \omega)$ at the Fermi level $\omega = 0$ when $\Delta^a=0$, with $\lambda=0$ for (c) and $\lambda\neq 0$ for (d). The yellow arrows in (b) and (d)indicates a Fermi pocket which may be probed by the quantum oscillation experiment.}
	\label{fig_pockets}
\end{figure}

Finally, we point out the existence of a new phase on top of the LPP with $\Delta^s\neq 0$ and $\langle h^{\dagger}\rangle \neq 0$. That is, the pairing of the $a$-spinons vanishes with $\Delta^a=0$. In this case, $|\Phi_a\rangle$ in Eq. (\ref{parent}) reduces to a Fermi liquid with the Fermi pockets shown in \figref{fig_pockets}, becoming gapless energy contours at $\omega=0$. Previously such a phase, known as the LPP-II \cite{Weng.Ma.2014} has been argued to be realized in a strong perpendicular magnetic field which suppresses $\Delta^a$ at least in the magnetic vortex core region. Consequently, the $a$-spinons become charged particles and respond to the detection of external electromagnetic fields in the LPP-II. It is also equivalently to say that the ``twisted'' quasiparticle $\tilde{c}$ becomes gapless forming the Fermi pockets in \figref{fig_pockets} since the holon is condensed in \eqnref{ctilde}.

In this $\Delta^a\rightarrow 0$ limit, the Landau's quasiparticle is no longer protected within the gapped $a$-pockets discussed in the main text, instead, they combine with gapless $a$-spinons to form lightly smaller Fermi pocket marked by yellow arrow in \figref{fig_pockets}(b) and (d). Note that the Fermi pocket for pure $a$-spinons shown in \figref{fig_pockets}(a) and (c) violates the transitional Luttinger sum rule, i.e., with area proportional to the hole concentration $\delta$, rather than $1+\delta$. Specifically, the area of each $a$-pocket shown in \figref{fig_pockets}(a) and (c) is given by
$A_{a}=\frac{\delta}{2} A_{\mathrm{BZ}}$,  where $A_{\mathrm{BZ}}=2\pi^2/a^2$ is the folding Brillouin zone due to the $\pi$-flux of $a$-spinons \cite{Weng.Ma.2014}. And the RPA correction gives rise to the Fermi pockets with a smaller area $A^\prime_a$ marked by yellow arrow in \figref{fig_pockets}(b) and (d), which can be measured by some experimental probes, e.g. quantum oscillation. According to the Onsager relation, $F=\frac{\Phi_{0}}{2 \pi^{2}} A^\prime_a$, where $\Phi_{0}$ is the magnetic flux quanta, giving the quantum oscillation $F=539.329 T$ at $\delta=0.1$, which is quite close to the experimental result $(530 \pm 20) T$ in Ref. \onlinecite{Taillefer.Doiron-Leyraud.2007} as compared to the original $F=697 T$ at the mean-field level \cite{Weng.Ma.2014} corresponding to \figref{fig_pockets}(c). It should be noted that the existence of zero poles along the large Fermi surface, indicated by the blue line in \figref{fig_pockets}(b), is not complete due to the diminished spectral weight in the anti-nodal regions, as identified by the gray region, resulting in the absence of a detectable quantum oscillation signal with an associated frequency.

Thus the pocket physics under strong magnetic fields may be regarded as a direct experimental probe of the ``twisted'' $\tilde c$ in the fractionalization \eqnref{Eq:1}, defined by \eqnref{ctilde}, which acquires small Fermi pockets at $\Delta^a=0$ as shown in \figref{fig_pockets}. The mechanism here is quite different from the previous proposals \cite{Lee.Senthil.2009, Sebastian.Harrison.2012, Zhang_2016}, where the small Fermi pocket is originated from the Fermi surface reconstruction due to a static CDW or spin order. In addition, together with the LPP, such an LPP-II state also disappears at $\delta>\delta^*$ where $\tilde c$ loses the coherence in the totally disordered spin background of $\Delta^s=0$.

\section{Conclusion}\label{sec7}

The quasiparticle is the most elementary excitation in a Fermi liquid. In contrast, the quasiparticle may further decay into more elementary fractionalized particles (i.e., bosonic $b$-spinons, fermionic backflow $a$-spinons and bosonic holons) in a strongly correlated electron system like doped Mott insulators. The fate of a quasiparticle excitation thus becomes an important manifestation of a non-Fermi-liquid state, even if it may no longer be the most essential constituent of the underlying phase.

In this paper, we have studied the single-particle Green's function in the doped Mott insulator described by the $t$-$J$ model, with the main focus on the evolution of the gapless quasiparticle excitation as a function of doping. The underlying phases under examination are low-temperature non-superconducting ones by removing the phase coherence. At low doping ($\delta<\delta^*$), it is known as the lower pseudogap phase or spontaneous vortex phase \cite{Weng.Ma.2014}, which is characterized by the two-component RVB orders. One is primary persisting to half-filling and the other is induced by doping. The primary RVB order parameter vanishes at $\delta^*$ to result in a strange-metal phase at overdoping ($\delta<\delta^*$). Such a phase diagram is regarded as the result of a mean-field \emph{parent} state given in Eq. (\ref{parent}) in the phase-string formulation of the $t$-$J$ model, in which the true long-range orders of antiferromagnetism and superconductivity can be further realized as the low-temperature instabilities \cite{Weng.Weng.2011,Weng.Ma.2014}.

The parent mean-field state in Eq. (\ref{parent}) has no quasiparticle excitations, which are fractionalized according to Eq. (\ref{Eq:1}). The quasiparticle excitation can emerge at a generalized RPA level as given by Eq. (\ref{Eq:7}) or Eq. (\ref{GLPG}), which is a diagrammatic Dyson equation expression of the following basic fractionalization process:
\begin{equation}\label{decay}
{c}_{i\sigma} \leftrightarrow \tilde{c}_{i\sigma}e^{i\hat{\Omega}_{i}}.
\end{equation}
Since the phase-shift factor $e^{i\hat{\Omega}_{i}}$ is disordered in the parent phase under examination to prevent the superconducting instability [cf. Eqs. (\ref{SC}) and (\ref{phdisor})], the quasiparticle and its fractionalized representation in Eq. (\ref{decay}) are two distinct excitations.
%as illustrated schematically in Fig . \ref{fig_ill}.
In the lower pseudogap phase, they represent the Fermi arc and the gapped incoherent portions, respectively. In the overdoping at $\delta>\delta^*$, $e^{i\hat{\Omega}_{i}}$ is totally disordered as $\Delta^s=0$ such that the propagation under the fractionalization in Eq. (\ref{decay}) is completely suppressed \emph{spatially}, which leads to a uniform reduction of the quasiparticle spectral weight in the momentum space. Consequently the full Fermi surface is explicitly present in the single-particle Green's function even though the spectral weight can be much reduced in the presence of an incoherent fractionalization in the background.

%in the underlying ground states in the absence of superconductivity on either side of the critical point, based on a single-particle Green’s function constructed in the phase string formulation of . The key of such phenomenological  quasiparticle propagator consisting of two components $D_{\tilde c}$ and $G_{C}$ in the general Dyson equation is that the bare holes lose all the coherence due to phase string sign structure, and thus can only emergent as the composite mode via recombinations of vortices and ``twisted'' holes. Here, in contrast to the bare injected hole propagator $G_{C}$ with a large Fermi surface satisfying the Luttinger volume of the total electrons, the leading term of emergent quasiparticles is the propagator of ``twisted'' holes $D_{\tilde c}$, which is essentially characterized by four small Fermi pockets of an area proportional to the doping concentration $\delta$ centered at $(\pm \pi/2, \pm \pi/2)$ as illustrated in \figref{fig_weight}.

%From the perspective of quasiparticles that can be measured by ARPES, the propagation of ``twisted'' hole $D_{\tilde c}$ is always accompanied by the vortices $e^{i\hat \Omega_i}$, so the behaviors of vortices in different phases highly determine the corresponding distinct behaviors of emergent quasiparticles. In the PG phase, the vortex remains relatively coherence so that the single-particle Green’s function in this work can give Fermi arc, of which
Therefore, the Fermi arc feature in the pseudogap phase is due to the fact that the quasiparticle weight is suppressed near the anti-nodal region, rather than a true energy gap is opened up in the quasiparticle excitation. On the other hand, the gap feature in the pseudogap phase actually comes from a fractionalized fermionic spinon in the single-particle Green's function, with the diminished quasiparticle spectral weight around the Fermi energy. The coherence of the gapless quasiparticle at the Fermi arc is protected by the Fermi pocket of the fermionic spinons, which are paired up to open an $s$-wave gap. A direct observation of the Fermi pocket may be realized by applying a strong magnetic field to suppress the spinon pairing, leading to a quantum oscillation as the fermionic spinons become charged at $\Delta^a=0$. Here, a small Fermi pocket picture violating the Luttinger volume does emerge for the ``twisted'' quasiparticle $\tilde{c}$. Furthermore, the endpoints of the Fermi arcs are marked by the ``hotspots'', which leads to a CDW-like quasiparticle interference pattern via scatterings between the ``hot-spots''. We have also discussed the consistency of these features with the experimental observations in the cuprate.

%When crossing the critical point, the divergent specific heat comes from the collapses of background $b$-spinons for the sake of the short enough AFM correlation length.
%%%%%%%%%%%%%%%%%%%%%%%%%%%%%%%%%%%%%%%%%%%%%%%%%%%%%%%%%%%%%%%%%%%%%%%%%%%%%%%%%%%%%%%%%%%%%%

\begin{acknowledgments}
    Stimulating discussions with Jian-Hao Zhang, Sen Li, Yao Ma, and Hong Ding are acknowledged.
    This work is partially supported by MOST of China (Grant No. 2017YFA0302902). \\
\end{acknowledgments}

%\section*{APPENDIX}

\appendix
%%%%%%%%%%%%%%%%%%%%%%%%%%%%%%%%%%%%%%%%%%%%%%%%%%%%%%%%%%%%%%%%%%%%%%%%%%%%%%%%%%%%%%%%%%%%%%
\section{Effective Hamiltonian}\label{app_mean-field}

The pseudogap ground state in \eqnref{parent} is a direct product of $\left|\Phi_{h}\right\rangle$, $\left|\Phi_{a}\right\rangle$, and $\left|\Phi_{b}\right\rangle$, which are the mean-field solutions of $H_{h}$, $H_{a}$, and $H_{b}$, respectively, of the effective Hamiltonian deduced from the phase-string formulation of the $t$-$J$ model \cite{Weng.Weng.2011, Weng.Ma.2014}:
\begin{equation}\label{eff}
	H_{\mathrm{eff}}=H_{h}+H_{a}+H_{b}
\end{equation}
with
\begin{eqnarray}
	H_{h} &=&-t_{h} \sum_{\langle i j\rangle} h_{i}^{\dagger} h_{j} e^{i\left( A_{ij}^{s}+eA_{ij}^e\right)}+h. c . \notag \\
	&\;&+\lambda_{h}\left(\sum_{i} h_{i}^{\dagger} h_{j}-\delta N\right) \label{Eq:Hh} \\
  H_{a} &=&-t_{a} \sum_{\langle i j\rangle, \sigma} a_{i \sigma}^{\dagger} a_{j \sigma} e^{-i \phi_{i j}^{0}} +h. c. -\gamma \sum_{\langle ij\rangle}(\hat{\Delta}_{i j}^{a})^{\dagger} \hat{\Delta}_{i j}^{a}\notag \\
   &\;& +\lambda_{a}\left(\sum_{i, \sigma} a_{i \sigma}^{\dagger} a_{i \sigma}-\delta N\right) \label{Eq:Ha}\\
   H_{b} &=&-J_{s} \sum_{\langle i j\rangle,\sigma} \hat{\Delta}_{i j}^{s}+h . c. +\lambda_{b}\left(\sum_{i, \sigma} b_{i \sigma}^{\dagger} b_{i \sigma}-N\right). \label{Eq:Hb}
\end{eqnarray}

Here in $H_h$, the holon $h_i^{\dagger}$ carries a full electric charge $+e$ coupling to the external electromagnetic field $A^e_{ij}$ as well as the internal gauge field defined by
\begin{equation}
	A_{i j}^{s}=\frac{1}{2} \sum_{l \neq i, j}\left[\theta_{i}(l)-\theta_{j}(l)\right]\left(n_{l \uparrow}^{b}-n_{l \downarrow}^{b}\right)
\end{equation}
which is generated by the background $b$-spinons. In the LPP, the fluctuations of this gauge field is suppressed, i.e., $A_{i j}^{s} \simeq 0$, because of the short-range RVB pairing of the $b$-spinons. Thus one expects the Bose condensation of the bosonic $h$-holons.

In $H_a$, the fermionic $a$-spinons see a uniform static $\pi$-flux per plaquette and form an $s$-wave BCS-type pairing $\Delta^{s}=\langle\hat{\Delta}_{i j}^{s}\rangle\neq0$, where
\begin{equation}
\hat{\Delta}_{i j}^{a}=\sum_{\sigma} \sigma a_{i \sigma}^{\dagger} a_{j \bar{\sigma}}^{\dagger} e^{-i \phi_{i j}^{0}},
\end{equation}
which is invariant under the gauge choice of the $\pi$-flux $\phi_{i j}^{0}$ \cite{Weng.Ma.2014}.

Inversely, in $H_b$ of \eqnref{Eq:Hb}, the bosonic RVB order-parameter operator is given by
\begin{equation}
	\hat{\Delta}_{i j}^{s}=\sum_{\sigma} e^{-i \sigma A_{i j}^{h}} b_{i \sigma} b_{j \bar{\sigma}}
\end{equation}
and the $b$-spinons couple with the internal gauge field generated by the holons:
\begin{equation}
 A_{i j}^{h}=\frac{1}{2} \sum_{l \neq i, j}\left[\theta_{i}(l)-\theta_{j}(l)\right] n_{l}^{h},
\end{equation}
which describes a uniform flux, on account of the holon condensation in the LPP. In \eqnref{Eq:Hb}, $J_s=J_{\text{eff}}\Delta^s/2$, $J_{\text{eff}}=J(1-\delta)^2-2\gamma\delta^2$, with $\Delta^{s}\equiv \langle\hat{\Delta}_{i j}^{s}\rangle\neq0$.

In the above effective Hamiltonians, $\lambda_h$, $\lambda_a$, and $\lambda_b$, are the Lagrangian multipliers implementing the constraints for the fractionalized particle numbers in \eqnref{projection}. And $\gamma$ in \eqnref{Eq:Ha} is the Lagrangian multiplier to enforce $\delta^{2}|\Delta^{s}|^2\approx \langle (\hat{\Delta}_{i j}^{a})^\dagger\hat{\Delta}_{i j}^{a}\rangle$, which is originated from the constrainit: $\boldsymbol{S}_{i}^{b} n_{i}^{h}+\boldsymbol{S}_{i}^{a}=0$ \cite{Weng.Ma.2014}. Note that $N$ denotes the total number of lattice sites.

\section{Mean-field self-consistent calculation}\label{sec_App_selfconsis}

Based on the mean-field Hamiltonian in \eqnref{Eq:Ha},  the free energy for the $a$-spinons is given by:
\begin{eqnarray}
F_{a}&=&-\frac{2}{\beta} \sum_{k,\alpha=\pm}{}^\prime  \ln \left[2 \cosh \left(\beta E_{k, \alpha}^a / 2\right)\right]\\
&\;&+\gamma 2 N\left(\left|\chi^{a}\right|^{2}+\left|\Delta^{a}\right|^{2}\right)+\lambda_{a} N(1-\delta)\notag
\end{eqnarray}
where $\sum_k^\prime$ denotes the summation over the half Brillouin zone due to the $\pi$-flux folding, and $E_{k,\pm}^a=\sqrt{(\xi_{k,\pm}^a)^2+\Delta_k^2}$ with $\xi_{k,\pm}^a=\pm2 (t_{a}+\gamma \chi^a) \sqrt{\cos ^{2} k_{x}+\cos ^{2} k_{y}}+\lambda_{a}$. Next, minimizing this mean-field free energy, i.e., $\partial F_{a} / \partial \chi^{a}=\partial F_{a} / \partial \Delta^{a}=\partial F_{a} / \partial \lambda_{a}=0$, gives rise to the self-consistent equations:
\begin{equation}\label{self1}
\begin{aligned}
\sum_{k, \alpha=\pm}{}^\prime& \gamma B_{k, \alpha} A_{k}= N \notag\\
\sum_{k, \alpha=\pm}{}^\prime& (-1)^{\alpha} \sqrt{A_{k}} B_{k, \alpha} \xi_{k, \alpha}^a=2 N \chi^{a}\notag\\
\sum_{k, \alpha=\pm}{}^\prime& \xi_{k, \alpha}^a B_{k, \alpha}=(1-\delta) N
\end{aligned}
\end{equation}
where $A_{k}=\cos ^{2} k_{x}+\cos ^{2} k_{y}$ and $B_{k, \alpha} = \tanh (\frac{1}{2} \beta E_{k, \alpha}^a) / E_{k, \alpha}^a$

Similarly, based on the mean-field Hamiltonian in \eqnref{Eq:Hb}, the free energy for the $b$-spinons is given by:
\begin{eqnarray}\label{freeb}
F_{b}&=&\frac{2}{\beta} \sum_{m} \ln \left(1-e^{-\beta E_{m}^b}\right)\notag\\
&\;&+\sum_{m} E_{m}^b+J_{\mathrm{eff}} |\Delta^{s}|^{2} N-2 \lambda_{b} N
\end{eqnarray}
where $E_m^b$ is the energy spectrum of $b$-spinons in \eqnref{Emb}, and $\partial F_{b} / \partial \Delta^{s}=\partial F_{b} / \partial \lambda_{b}=0$ give rise to the self-consistent equations:
\begin{eqnarray}\label{self2}
&\;&\sum_{m} \frac{(\xi_{m}^{b})^2 \operatorname{coth}\left(\frac{1}{2} \beta E_{m}^b\right)}{E_{m}^b}-2 N (\Delta^{s})^{2} J_{\mathrm{eff}}=0 \notag\\
&\;&\sum_{m} \frac{\lambda_{b} \operatorname{coth}\left(\frac{1}{2} \beta E_{m}^b\right)}{E_{m}^b}=2 N
\end{eqnarray}
where $\xi_m^b$ is determined by the self-consistent equation in \eqnref{Hbdiag}. Since the bosonic holons are in the condensation state in the LPP, their contribution to the free energy may be neglected here such that by minimizing the total free $F_a+F_b$ over $\gamma$, one obtains
\begin{equation}\label{self3}
	\delta^{2} |\Delta^{s}|^{2}=|\Delta^{a}|^{2}+4 \chi_{a}^{2}
\end{equation}
The values of the parameters used in the main text, i.e., $\chi_a$, $\Delta^a$, $\lambda_a$, $\Delta^s$, $\lambda_b$, and $\gamma$, at different doping concentrations $\delta$, are determined by the self-consistent calculations based on the above equations in \eqnref{self1}, \eqnref{self2} and \eqnref{self3}, under the choice of $t_a= 2J$ which is the same as in Ref. \onlinecite{Weng.Ma.2014}.
%%%%%%%%%%%%%%%%%%%%%%%%%%%%%%%%%%%%%%%%%%%%%%%%%%%%%%%%%%%%%%%%%%%%%%%%%%%%%%%%%%%%%%%%%%%%%%
\section{Phase factor $\exp \left\{-\frac{i}{2}\left(\Phi_{i}^{0}-\Phi_{j}^{0}\right)\right\}$ in \eqnref{Eq:3}} \label{sec:simplify}

By inserting a sequence of nearest-neighboring links connecting site $i$ and site $j$ (denoting $i \rightarrow j$): $i_1$, $i_2$ ... $i_M$, then one has
\begin{equation}
\begin{aligned}\label{ph}
    \exp &\left\{-\frac{i}{2}(\Phi_{i}^{0}-\Phi_{j}^{0})\right\}\\
	&=\exp \left\{-\frac{i}{2}(\Phi_{i}^{0}-\Phi_{i_{1}}^{0}+\Phi_{i_{1}}^{0}-\Phi_{i_{2}}^{0}+\ldots+\Phi_{i_{M}}^{0}-\Phi_{j}^{0})\right\} \\
    &=\exp [ -i \sum_{i \rightarrow j} \phi_{i_{s}, i_{s+1}}^{0}]\\
	&\;\;\;\;\;\times \prod_{i \rightarrow j} \exp \left\{-\frac{i}{2}\left[\theta_{i_{s}}(i_{s+1})-\theta_{i_{s+1}}(i_{s})\right]\right\}
\end{aligned}
\end{equation}
where $\phi_{i_{s}, i_{s+1}}^{0}=\frac{1}{2} \sum_{l \neq i_{s}, i_{s+1}}\left[\theta_{i_{s}}(l)-\theta_{i_{s+1}}(l)\right]$. Note that $\theta_{i_{s}}(i_{s+1})-\theta_{i_{s+1}}(i_{s})=\pm\pi$,  the second term in \eqnref{ph} can be further simplified as:
\begin{equation}
\begin{aligned}\label{simph}
  &\prod_{i \rightarrow j} \exp  \left\{-\frac{i}{2}\left[\theta_{i_{s}}(i_{s+1})-\theta_{i_{s+1}}(i_{s})\right]\right\}\\
  &=\left(e^{\pm i\pi/2}\right)^{i-j}=e^{i \boldsymbol{k}_0 \cdot(\boldsymbol{r}_i-\boldsymbol{r}_j)}
\end{aligned}
\end{equation}
where $\boldsymbol{k}_{0}=(\pm\pi / 2, \pm\pi / 2)$. Note that, although there exists a freedom to choose $\pm \pi$ at each link since $\exp\left\{-\frac{i}{2} \theta_{i_{s}}(i_{s+1})\right\}$ here is multi-valued, all the branch-cuts in \eqnref{simph} are chosen the same to make the phase factor ``smooth''. As the result, one arrives at
\begin{equation}\label{reph}
  \exp \left\{-\frac{i}{2}\left(\Phi_{i}^{0}-\Phi_{j}^{0}\right)\right\}=\exp[-i \sum_{i \rightarrow j} \phi_{i_{s}, i_{s+1}}^{0}]\times e^{i \boldsymbol{k}_{0} \cdot\left(\boldsymbol{r}_{i}-\boldsymbol{r}_{j}\right)}
\end{equation}

\section{Single-particle Green's function in the SC phase}\label{sec_App_SC}

\begin{figure}[t]
	\centering
	\includegraphics[scale=0.9]{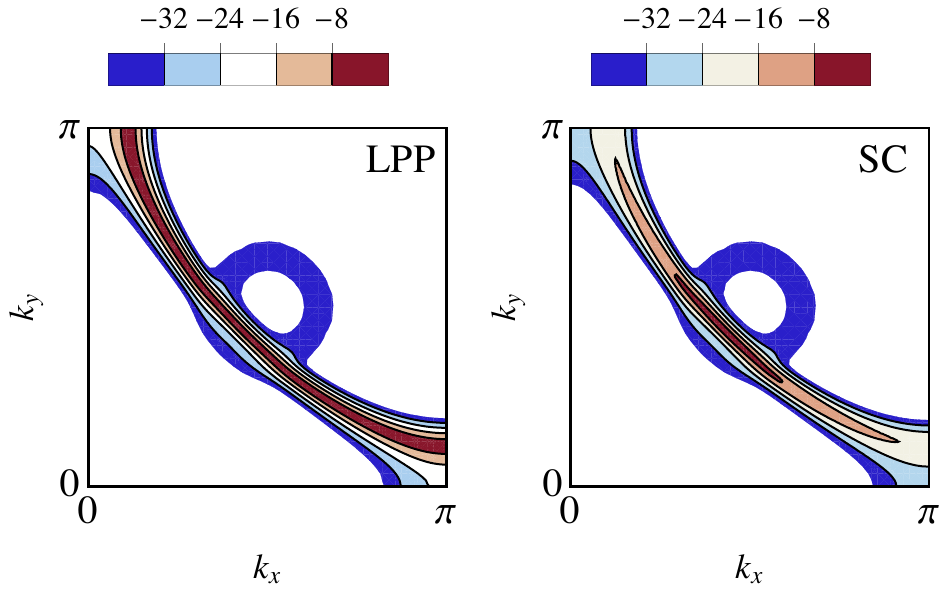}
	\caption{The contours of the poles of the single-particle Green's function near Fermi surface in the LPP (left panel) and the SC phase (right panel). The doping concentration is at $\delta = 0.1$.
	}
	\label{fig_SC}
\end{figure}

As the phase coherence of the phase factor is realized, i.e., $\langle e^{i \Phi_{i}^{s}(\tau)}\rangle \neq 0$, the SC coherence is established in Eq. (\ref{SC}), and the Nambu representation for the Green's function will be needed, with $\Psi_a=[\begin{array}{cc}a_{k \uparrow}^{\dagger} & a_{-k \downarrow}\end{array}]^{T}$ and $\Psi_c=[\begin{array}{cc}c_{k\uparrow}  & c_{-k \downarrow}^{\dagger}\end{array}]^{T}$. Namely, the bare propagators should be extended to:
\begin{equation}
\begin{aligned}\label{fSC}
  \mathcal{D}_{a}(\boldsymbol{k} ; \omega) &\rightarrow \frac{1}{\omega \tau_{0}+\xi_{k}^{a} \tau_{z}-\Delta_{k} \tau_{x}} \\
	G_{0}^{c}(\boldsymbol{k} ; \omega) &\rightarrow \frac{1}{\omega \tau_{0}-\epsilon_{k}^{c} \tau_{z}-\Delta_{k}^{c} \tau_{y}}
\end{aligned}
\end{equation}
where $\Delta_{k}^{c}=J_{\mathrm{eff}} \Delta^{a}\left(\cos k_{x}-\cos k_{y}\right)$ is the $d$-wave pairing order parameter. In addition, the phase can also be rewritten as $e^{\frac{i}{2} \Phi_{i}^{s}(\tau)} \rightarrow [\begin{array}{cc}e^{\frac{i}{2} \Phi_{i}^{s}(\tau)}  & e^{-\frac{i}{2} \Phi_{i}^{s}(\tau)}\end{array}]^{T}$
, so the phase correlation function in \eqnref{f} is modified as:
\begin{equation}\label{ffSC}
	f\left(r_{i}-r_{j}, \tau\right)\approx F_{0} \tau_{0}+F_{0}\left\langle e^{i \Phi_{i}^{s}}\right\rangle_{0} \tau_{x}
\end{equation}
As the result, according to the Dyson equation \eqnref{Eq:6}, the single-particle Green's function in the SC phase is given by:
\begin{equation}\label{tSC}
	G^{e}_{\text{SC}}(\boldsymbol{k} ; \omega) = \frac{\boldsymbol{M} \cdot \boldsymbol{\tau}}{\left(\omega^{2}-E_{+}^{2}\right)\left(\omega^{2}-E_{-}^{2}\right)}
\end{equation}
where $\boldsymbol{\tau}\equiv [\begin{array}{cccc}\tau_0 & \tau_x & \tau_y & \tau_z \end{array}]$ as well as $E_{\pm}=\sqrt{A / 2 \pm \sqrt{B} / 2}$, with $A=(\Delta_{k+k_{0}}^{a})^{2}+(\Delta_{k}^{c})^{2}+(\xi_{k+k_{0}}^{a})^{2}+(\epsilon_{k}^{c})^{2}+2 \lambda^{2}$ and $B=((\Delta_{k+k_{0}}^{a})^{2}-(\Delta_{k}^{c})^{2}+(\xi_{k+k_{0}}^{a})^{2}-(\epsilon_{k}^{c})^{2})^{2}+4((\Delta_{k+k_{0}}^{a})^{2}+(\Delta_{k}^{c})^{2}+(\xi_{k+k_{0}}^{a}-\epsilon_{k}^{c})^{2}) \lambda^{2}$. Although \eqnref{tSC} is different from the Green's function in traditional BCS theory, its low-energy branch of the poles $E_{-}$ in the SC phase still open a $d$-wave-like gap in the nodal region as shown in the right panel of \figref{fig_SC} (banana-shaped contours), which is contrary to the gapless case in the LPP shown in the left panel of \figref{fig_SC}.

Finally, note that as the SC phase coherence is disordered, i.e., $\langle e^{i \Phi_{i}^{s}(\tau)}\rangle = 0$, the anomalous part for $G_{0}^{e}(\boldsymbol{k} ; \omega)$ vanishes since the phase correlation function in \eqnref{ffSC} reduces to an identity matrix. Furthermore, the anomalous part for $G_{0}^{c}(\boldsymbol{k} ; \omega)$ also disappears as $\Delta_{k}^{c}=0$. Therefore, the single-particle Green's function in the LPP discussed in the main text can be smoothly connected to that in the SC phase.

\section{Origin of the symmetric peak in the QPI pattern}\label{sec_App_QPI}
\begin{figure}[t]
	\centering
	\includegraphics[scale=0.55]{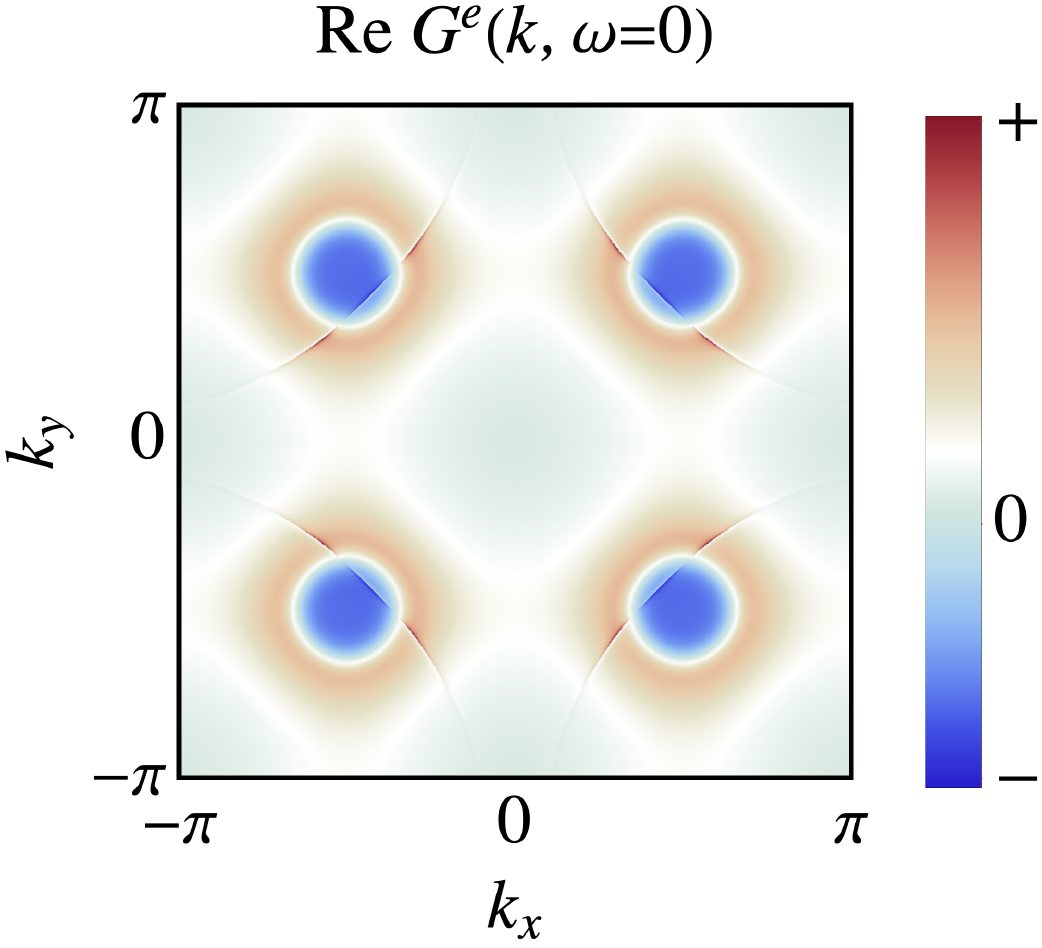}
	\caption{$\operatorname{Re} G^e(\boldsymbol k, \omega=0)$ at $\delta=0.1$, here the lifetime of bare hole takes the value of $\eta=1 \text{meV}$ when taking $\omega \rightarrow \omega+i\eta$ in \eqnref{GLPG}. Background weights seem give nested structure, but the narrow contours along the large Fermi surface violate the nesting condition.
	}
	\label{fig_app1}
\end{figure}

To understand the origin of the symmetric QPI pattern in \figref{fig_CDW}(b), we expand \eqnref{LDOS} to the leading term in $v_i$:
\begin{equation}
	\delta\rho(\boldsymbol{q},\omega)\simeq -\frac{1}{\pi}  v(\boldsymbol{q}) \operatorname{Im} \Lambda(\boldsymbol{q}, \omega)
\end{equation}
where
\begin{equation}
	\Lambda(\boldsymbol{q}, \omega)=\sum_{\boldsymbol{k}}G^e(\boldsymbol{q}, \omega)  G^e(\boldsymbol{k}-\boldsymbol{q}, \omega)
\end{equation}
Here $v(\boldsymbol{q})= \sum_{i} v_{i} e^{-i\boldsymbol{q} \cdot \boldsymbol{r}_{i}}$ is a random function of $\boldsymbol{q}$ when the number of impurities is sufficient, thus all the intrinsic information about quasiparticles is encoded in the $\operatorname{Im}\Lambda(\boldsymbol{q}, \omega)$, which can be further decomposed as:
\begin{equation}\label{QPIFS}
		\operatorname{Im}\Lambda(\boldsymbol{q}, \omega) \varpropto  \sum_{\boldsymbol{k}}\operatorname{Re} G^e(\boldsymbol{q}, \omega) \operatorname{Im}G^e(\boldsymbol{k}-\boldsymbol{q}, \omega)
\end{equation}
Note that $\operatorname{Im} G^e$ is actually corresponding to the spectral function illustrated in \figref{fig:arcevo}, but $\operatorname{Re} G^e$ gives pseudo-nesting structure with nesting vector $\boldsymbol{q}_0=(\pi,0)$ and $(0,\pi)$ at zero energy, i.e., $\operatorname{Re} G^e(\boldsymbol{q}, \omega=0) \simeq \operatorname{Re} G^e(\boldsymbol{q}-\boldsymbol{q}_0, \omega=0)$, which is shown in \figref{fig_app1}. More specifically, the background weights that contribute most to nested structure in \figref{fig_app1} come from the perfect nesting condition for ``twisted'' hole $\tilde{c}_i$ as shown in \figref{a-real}(b), while the narrow contours along the large Fermi surface from bare holes lightly break nesting condition (reason for ``pseudo'' here). At the current setting parameters, the former dominates since the ``twisted'' hole propagator is the leading term for the single-particle Green's function in \eqnref{GPG}. Therefore, it is the pseudo-nested structure of $\operatorname{Re} G^e$ results in $\delta\rho(\boldsymbol{q},\omega=0)\simeq\delta\rho(\boldsymbol{q}-\boldsymbol{q}_0,\omega=0)$. Combined with the relation $\delta\rho(\boldsymbol{q},\omega=0)\simeq\delta\rho(-\boldsymbol{q},\omega=0)$, one finally arrives at:
\begin{equation}
	\delta\rho(\boldsymbol{q},\omega=0)\simeq\delta\rho(\boldsymbol{q}_0-\boldsymbol{q},\omega=0)
\end{equation}
which explains the cause of two symmetric peaks about $(\pi/2,0)$ in QPI pattern \figref{fig_CDW}(b).

In summary, the peak labeled by the dashed line in \figref{fig_CDW}(b) origins from the scattering between two parallel neighboring ``hotspots'', while the  peak labeled by the arrow comes from the special symmetrical structure of $\tilde{c}_i$ spectrum. The latter may be not intrinsic since we can remove its perfect nesting condition by either introducing more parameters and higher-order interaction to change its pocket shape, or tuning parameter, such as coupling constants $\lambda$ and bare hole life-time $i\eta$, to enhance the bare hole contribution in $G^e$.

\section{Dynamical charge susceptibility}

	In this section, we aim to investigate the potential charge instabilities captured by the dynamical charge susceptibility within our framework and establish the connection between the QPI pattern presented in the main text and the dynamical charge susceptibility calculated in this section.

\begin{figure}[hptb]
	\centering
	\includegraphics[scale=0.35]{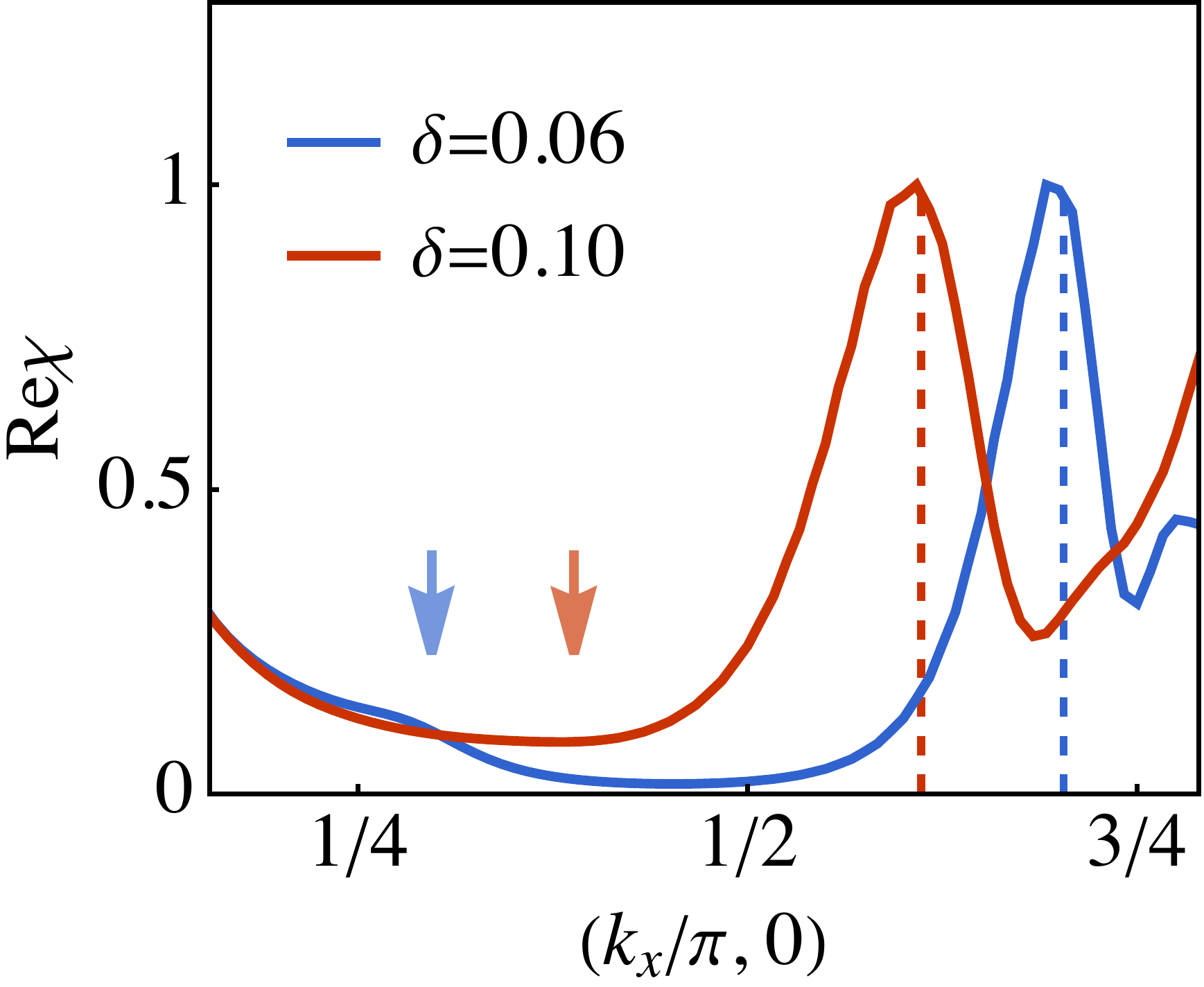}
  \caption{The real part of the free charge susceptibility $\Re\chi$ (renormalized by dividing by the maximum value) is shown for different doping densities. The dashed lines correspond to the peaks, with momenta aligning with $Q_0$ connecting two parallel hotspots, as expounded in the main text. The symmetric redundancy peak location, mentioned in the main text, is identified using arrows indicating the momentum.}
	\label{fig:CDW}
\end{figure}

	The standard dynamical charge susceptibility, denoted as $\chi_c$, is defined with 
	\begin{equation}
		\chi_c(i, j, \tau) \equiv\left\langle\hat{T} \hat{n}_j(\tau) \hat{n}_i(0)\right\rangle_0,
	\end{equation}
	where $\hat{n}_i=\sum_\sigma \sigma c_{i \sigma}^{\dagger}(\tau) c_{i \sigma}(\tau)$ is the number of electrons operator. By expanding in Wick's theorem and Fourier transforming, we can express $\chi_c$ with:
	\begin{equation}
    	\chi_c\left(\boldsymbol{q}, i \omega_n\right)=-\frac{2}{\beta N} \sum_{\boldsymbol{k}, v_n}  G_{c^*}\left(\boldsymbol{k}, i v_n\right) G_c\left(\boldsymbol{q}-\boldsymbol{k}, i \omega_n-i v_n\right)
	\end{equation}
	where $G_c$ and $G_{c^*}$ denote the single-particle Green's function for quasiparticles and quasiholes, respectively. Further, using the spectral expansion with
	\begin{equation}
    	G_c\left(\boldsymbol{k}, i \omega_n\right)=\int_{-\infty}^{+\infty} \frac{d \Omega}{\pi} \frac{A_c(\boldsymbol{k}, \Omega)}{i \omega_n-\Omega},
	\end{equation}
	where $A_c(k, \omega)=-\operatorname{Im} G_c(k, \omega)$ represents the spectral function, and then applying Masubara summation and analytic continuation, we find that, at the mean-field level (single bubble diagram), the charge spectrum tends to zero, i.e.,
	\begin{equation}
    	\text{Im} \chi_c(\boldsymbol{q}, \omega=0)=0,
	\end{equation}
	which might suggest that there is no static charge order. However, it is important to note that the potential to induce CDW instability is actually hidden in $\chi_c$. We clarify this by examining the real part of $\chi_c$:
\begin{eqnarray}
    \Re \chi_c(\boldsymbol{q}, &\omega&=0) \notag \\
	&\propto& \int_0^{\Lambda} d \Omega \int_0^{\Lambda} d \Omega^{\prime} \sum_{\boldsymbol{k}} A_c(\boldsymbol{k},-\Omega)  A_c\left(\boldsymbol{q}-\boldsymbol{k}, \Omega^{\prime}\right) \notag \\
	&& \;\;\;\;\;\;\;\;\;\;\;\;\;\;\;\;\;\;\;\;\;\;\;\;\;\;\;\;\;\; \times \frac{1}{\Omega+\Omega^{\prime}}\label{rechi} \\
    &\propto&\sum_{\boldsymbol{k}} A_c(\boldsymbol{k},\Omega=0) A_c\left(\boldsymbol{q}-\boldsymbol{k}, \Omega^{\prime}=0\right),\label{reschi}
\end{eqnarray}
where \eqnref{reschi} is obtained by approximating that the main contribution to $\Re \chi_c$ arises from frequencies near zero (i.e. $\Omega=\Omega^\prime=0$) due to the presence of $1/(\Omega+\Omega^{\prime})$ in \eqnref{rechi}. \eqnref{reschi} indicates that the real part of the charge susceptibility at zero frequency is the result of the convolution of the spectral function at the Fermi surface. This suggests that regions with high density at the Fermi surface will make a significant contribution to the value of $\Re \chi_c$. The calculation result of $\Re \chi_c$ is presented in \figref{fig:CDW}, which shows a peak at the same momentum $Q_0$ (indicated by dashed lines) as discussed in \secref{secCDW}.

Next, assuming the presence of residual interactions with a strength of $g$ that can decompose the charge channel, the correction at the Random Phase Approximation (RPA) level yields 
\begin{equation}
    \chi_c^{\text{RPA}}(\boldsymbol{q})=\frac{\chi_c(\boldsymbol{q})}{1-g \chi_c(\boldsymbol{q})},\label{chicRPA}
\end{equation} 
which shows that the charge susceptibility $\chi_c^{\text{RPA}}$ can be enhanced at the momentum $Q_0$ due to the smallest value of $1-g \chi_c(\boldsymbol{q})$ in the denominator of \eqnref{chicRPA} at this particular momentum.

In addition, the charge susceptibility results reveal the disappearance of additional peaks (marked by arrows in the QPI mode in \figref{fig_CDW}(b)) in \figref{fig:CDW}. This suggests that these redundant peaks are not intrinsic and is consistent with the argument presented in Appendix E.

In summary, in the main text and Appendix E, we have demonstrated that the QPI pattern can reveal information about the momentum distribution of density at the Fermi surface (as described in \eqnref{QPIFS}). The scattering between the hotspots located at the ends of the Fermi arcs leads to a charge modulation pattern. Additionally, this section shows that any feature of the quasiparticle spectrum at the Fermi surface can be represented in the dynamical charge susceptibility at the mean-field level. This is similar to the case of a free Fermi liquid, where the "hot spots" or "nesting" structure at the Fermi surface can lead to peaks at corresponding momenta. However, in the free theory, the peak of the charge susceptibility does not diverge unless the RPA correction is considered for a specific channel. This implies that the leading order calculation cannot produce direct static charge order ($\langle n_q \rangle =0$) without interactions with specific charge channels. In our work, we avoid discussing the specific form of interaction and instead utilize impurities to mimic the effect of the interaction. The $t$-matrix method can then be used to directly obtain the finite static charge order. In other words, the impurities can act as a mixture of channels, including the charge channel.

%First, under the $t$-matrix method, the quasiparticle interference pattern can reveal the features of the momentum distribution of density at the Fermi surface (details are shown in the Appendix. E). Scattering between regions with high density induces peaks of quasiparticle interference intensity, which correspond to the scatthering between hopspots shown in the main text of our paper. Next, we will present the instability of the charge density wave calculated from the dynamical charge susceptibility $\chi_c\left(\boldsymbol{q}, i \omega_n\right)$.

%The standard dynamical charge susceptibility $\chi_c$ also depends on the momentum distribution of density at the Fermi surface.

\section{Dynamic spin susceptibility of $b$-spinons}\label{spin}

According to the mean-field Hamiltonian \eqnref{Eq:Hb}, the $b$-spinons perceive the uniform static gauge field with $\delta\pi$ flux per plaquette via $A_{ij}^h$ generated by holons in the condensate states. Thus, with the standard diagonalization procedure as in Hofstadter system, we obtain:

\begin{equation}
	H_{b}=\sum_{m, \sigma} E_{m}^b \gamma_{m \sigma}^{\dagger} \gamma_{m \sigma}
\end{equation}
with the $b$-spinons spectrum:
\begin{equation}\label{Emb}
	E_{m}^b=\sqrt{\lambda_b^{2}-(\xi_{m}^b)^{2}}
\end{equation}
via introducing the following Bogoliubov transformation:
\begin{equation}\label{bogo}
	b_{i \sigma}=\sum_{m} \omega_{m \sigma}(i)\left(u_{m}^b \gamma_{m \sigma}-v_{m}^b \gamma_{m-\sigma}^{\dagger}\right).
\end{equation}

\begin{figure}[b]
	\includegraphics[scale=0.7]{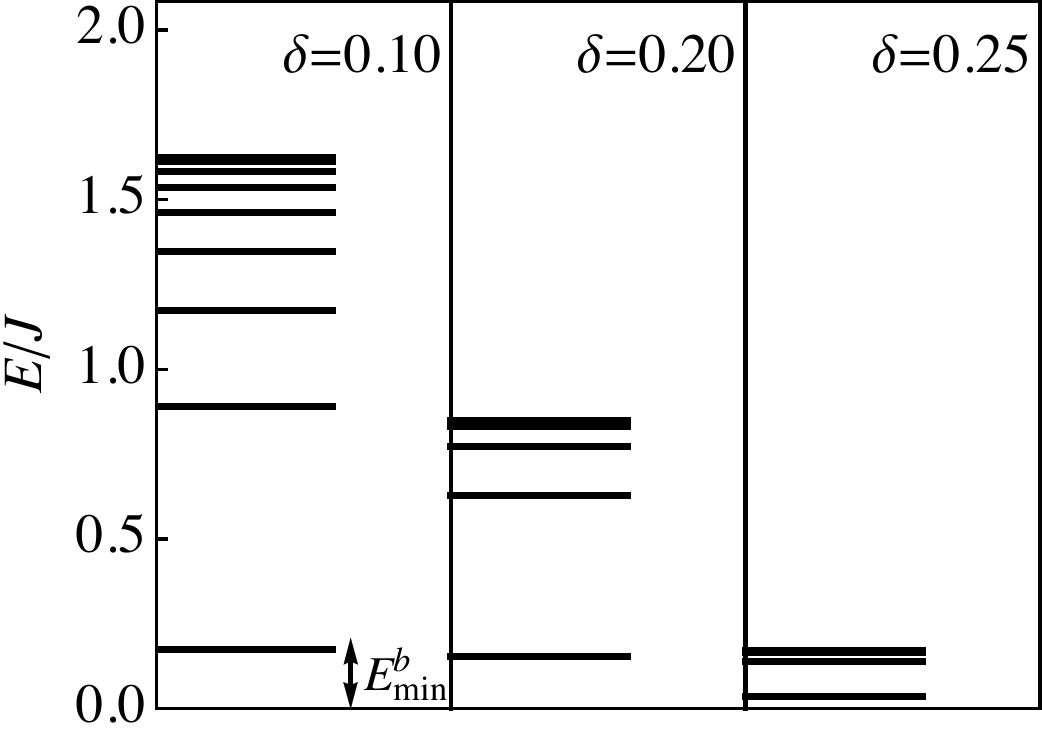}
	\caption{The evolution of $b$-spinon energy levels with respect to doping concentration when $\delta<\delta^*$ based on the mean-field self-consistent calculations. The lowest excitation energy gap is labeled by $E_{\text{min}}^b$ here.
	}
	\label{fig_spin}
\end{figure}

Here, the coherent factors are given by
\begin{eqnarray}
u_{m}^b&=&\sqrt{\frac{1}{2}\left(1+\frac{\lambda}{E_{m}^b}\right)}
\notag\\
	v_{m}^b&=&\operatorname{sgn}\left(\xi_{m}^b\right) \sqrt{\frac{1}{2}\left(-1+\frac{\lambda}{E_{m}^b}\right)}
\end{eqnarray}
and $\xi_{m}^b$ as well as $w_{m}(i)=w_{m \sigma}(i)=w_{m-\sigma}^{*}(i)$ are the eigenfunctions and eigenvalues of the following equation:
\begin{equation}\label{Hbdiag}
	\xi_{m}^b \omega_{m }(i)=-\frac{J \Delta^{s}}{2} \sum_{j=\text{NN}(i)} e^{i \sigma A_{i j}^{h} }\omega_{m }(j)
\end{equation}
where $J_s=J_{\text{eff}}\Delta^s/2$ and $J_{\text{eff}}=J(1-\delta)^2-2\gamma\delta^2$.

In this scheme, the parameters $\lambda_b$ and $\Delta^s$ are determined by self-consistent calculation \eqnref{self2}. The energy level for $b$-spinons $E_m^b$ at different doping is shown in \figref{fig_spin}, which is composed of the  ``Landau-level-like" discrete levels with a finite energy gap.

Then, by the relation $S_{i}^{b, z}=\frac{1}{2} \sum_{\sigma} \sigma b_{i \sigma}^{\dagger}b_{i \sigma}$, the Matsubara spin-spin correlation function can be expressed as:
\begin{eqnarray}
&\;&\chi^{zz}(\boldsymbol{r}_i- \boldsymbol{r}_j, \tau) =\left\langle\hat{T} S_{j}^{b, z}(\tau) S_{i}^{b, z}(0)\right\rangle_{0}
\\
	&=&\frac{1}{4} \sum_{\sigma \sigma^{\prime}} \sigma \sigma^{\prime}\left\langle\hat{T} b_{j \sigma}^{\dagger}(\tau) b_{j \sigma}(\tau) b_{i \sigma^{\prime}}^{\dagger}(0) b_{i \sigma^{\prime}}(0)\right\rangle_{0}\notag \\
	&=&\frac{1}{4} \sum_{\sigma \sigma^{\prime}} \sigma \sigma^{\prime}\left[\left\langle\hat{T} b_{j \sigma}^{\dagger}(\tau) b_{i \sigma^{\prime}}^{\dagger}(0)\right\rangle_{0}\left\langle\hat{T} b_{j \sigma}(\tau) b_{i \sigma^{\prime}}(0)\right\rangle_{0}\right. \notag \\
	&\;&\left.+\left\langle\hat{T} b_{j \sigma}^{\dagger}(\tau) b_{i \sigma^{\prime}}(0)\right\rangle_{0}\left\langle\hat{T} b_{j \sigma}(\tau) b_{i \sigma^{\prime}}^{\dagger}(0)\right\rangle_{0}\right]
\end{eqnarray}
where $\langle\rangle_0$ denotes the expectation value under the mean-field state, and the Wick's theorem is applied in the last line. Then, by using the Bogoliubov transformation \eqnref{bogo}, together with the Green's function
\begin{eqnarray}
	G_{\gamma}\left(m, i \omega_{n} ; \sigma\right) &\equiv& -\left\langle\gamma_{m \sigma}(i\omega_n) \gamma_{m \sigma}^{\dagger}(i\omega_n)\right\rangle_{0}\notag\\
	&=&\frac{1}{i \omega_{n}-E_{m}^b}
\end{eqnarray}
the spin correlation function in momentum and frequency space at $T=0$ can be expressed as:
\begin{eqnarray}
	\chi^{zz}\left(\boldsymbol{Q}, i \omega_{n}\right)&=&\frac{1}{8} \sum_{m m^{\prime}} C_{m m^{\prime}}(\boldsymbol{Q})\left(\frac{\lambda^{2}-\xi_{m}^b \xi_{m^{\prime}}^b}{E_{m}^b E_{m^{\prime}}^b}-1\right) \\
	&\;& \times \left(\frac{1}{i \omega_{n}+E_{m}^b+E_{m^{\prime}}^b}-\frac{1}{i \omega_{n}-E_{m}^b-E_{m^{\prime}}^b}\right)\notag
\end{eqnarray}
where
\begin{equation}
	C_{m m^{\prime}}(\boldsymbol{Q}) \equiv \sum_{i j} \frac{1}{N} e^{-i \boldsymbol{Q} \cdot\left(\boldsymbol{r}_{i}-\boldsymbol{r}_{j}\right)} w_{m}^{*}(i) w_{m^{\prime}}^{*}(j) w_{m^{\prime}}(i) w_{m}(j)
\end{equation}
Finally, the imaginary part of dynamic spin susceptibility can be obtained by the analytic continuation $\chi^{\prime \prime}(\boldsymbol{Q}, \omega)=\left.\operatorname{Im} \chi^{\mathrm{zz}}\left(\boldsymbol{Q}, i \omega_{n}\right)\right|_{i \omega_{n} \rightarrow \omega+i 0^{+}}$, resulting:
\begin{eqnarray}
	\chi^{\prime \prime}(\boldsymbol{Q}, \omega)&=&\frac{\pi}{8} \sum_{m m^{\prime}} C_{m m^{\prime}}(\boldsymbol{Q})\left(\frac{\lambda^{2}-\xi_{m}^b \xi_{m^{\prime}}^b}{E_{m}^b E_{m^{\prime}}^b}-1\right)\notag\\
	&\;&\times \operatorname{sgn}(\omega) \delta\left(|\omega|-E_{m}^b-E_{m^{\prime}}^b\right)\label{chib}
\end{eqnarray}
%%%%%%%%%%%%%%%%%%%%%%%%%%%%%%%%%%%%%%%%%%%%%%%
\section{Specific heat for $b$-spinons}\label{App_Cvb}
The contribution to the specific heat from $b$-pinons can be expressed as:
\begin{equation}
	\gamma \equiv \frac{C_{V}^b}{T}=-\frac{1}{N} \frac{\partial^{2}}{\partial T^{2}} F_{b}
\end{equation}
where $F_{b}$ is the free energy for $b$-spinons given in \eqnref{freeb}. Thus, the specific heat can be further written as:
\begin{equation}\label{eqnCv}
	\gamma = \frac{1}{N} \sum_{m} \frac{2 (E_{m}^b)^{2}}{k_{B} T^{3}} n_{B}(E_{m}^b)\left[n_{B}(E_{m}^b)+1\right]
\end{equation}
where $n_{B}(\omega)=1 /(e^{\beta \omega}-1)$ is the Bose distribution function and $E_{m}^b$ is the $b$-spinons spectrum given in \eqnref{Emb}.
\bibliography{revision_CP.bib}
%\bibliography{refs/draft_CP.bib}
\end{document}